\documentclass[twoside,leqno,twocolumn]{article}

\usepackage[letterpaper]{geometry}

\usepackage{siamproceedings}

\usepackage[T1]{fontenc}
\usepackage{amsfonts}
\usepackage{graphicx}
\usepackage{epstopdf}
\usepackage{enumitem}
\usepackage{algorithmic}
\ifpdf
  \DeclareGraphicsExtensions{.eps,.pdf,.png,.jpg}
\else
  \DeclareGraphicsExtensions{.eps}
\fi

\newsiamremark{remark}{Remark}
\newsiamremark{hypothesis}{Hypothesis}
\crefname{hypothesis}{Hypothesis}{Hypotheses}
\newsiamthm{claim}{Claim}

\usepackage{amsmath,amssymb,amsfonts}
\usepackage{algorithm}
\usepackage{algorithmic}
\usepackage[algo2e, noend,vlined]{algorithm2e}
\SetAlCapFnt{\small}    % caption font
\SetAlFnt{\small}       % algorithm font
\SetNlSty{}{}{:}        % line‐number style
\SetAlgoNlRelativeSize{-4}% slightly smaller line‐numbers
\usepackage{graphicx}
\usepackage{textcomp}
\usepackage{xcolor}
\usepackage{framed}
\usepackage{balance}
\usepackage{subcaption}
\usepackage{enumitem} 
\usepackage{booktabs}
\usepackage{multirow}
\usepackage{enumitem}

\setlist{nosep}
\usepackage{wrapfig}
\usepackage{threeparttable}

\definecolor{my_color}{rgb}{.30,.35,.50}

\definecolor{my_color2}{rgb}{.20,.35,.30}

\definecolor{my_color3}{rgb}{.90,.35,.30}

\begin{document}

\newcommand\relatedversion{}
\renewcommand\relatedversion{\thanks{The full version of the paper can be accessed at \protect\url{https://arxiv.org/abs/0000.00000}}} % Replace URL with link to full paper or comment out this line

\title{Mapping Sparse Triangular Solves to GPUs \\ via Fine-grained Domain Decomposition}
\author{%
  Atharva Gondhalekar\thanks{%
    Dept.\ of Electrical and Computer Engineering, Virginia Tech, Blacksburg, VA, USA
    (\email{atharva1@vt.edu}).%
  }%
  \and
  Kjetil Haugen\thanks{%
    Haugen Labs,
    (\email{kjteil@haugenlabs.com}).%
  }%
  \and
  Thomas Gibson\thanks{%
    Advanced Micro Devices, Inc., Austin, TX, USA
    (\email{thomas.gibson@amd.com}).%
  }%
  \and
  Wu-chun Feng\thanks{%
    Dept.\ of Computer Science, Virginia Tech, Blacksburg, VA, USA
    (\email{wfeng@vt.edu}).%
  }%
}

% \author{Atharva Gondhalekar\footnotemark[1]
      
%     \and Kjetil Haugen\footnotemark[2]
        
%     \and Thomas Gibson\footnotemark[3]
%     \and Wu-chun Feng\footnotemark[4] 
    
%     }

%\thanks{Dept. of CS \& ECE, Virginia Tech, Blacksburg, VA, USA (\email{wfeng@vt.edu})}

%\thanks{Advanced Micro Devices Inc., Austin, TX, USA (\email{thomas.gibson@amd.com})}
\date{}

\maketitle
\begin{abstract}

Sparse linear systems are typically solved using preconditioned iterative methods, but applying preconditioners via sparse triangular solves introduces bottlenecks due to irregular memory accesses and data dependencies.
This work leverages fine-grained domain decomposition to adapt triangular solves to the GPU architecture. %, optimizing the performance of preconditioned iterative solvers.
We develop a fine-grained domain decomposition strategy that generates non-overlapping subdomains, increasing parallelism in the application of preconditioner at the expense of a modest increase in the iteration count for convergence.
Each subdomain is assigned to a thread block and is sized such that the subdomain vector fits in the GPU shared memory, eliminating the need for inter-block synchronization and reducing irregular global memory accesses.

Compared to other state-of-the-art implementations using the ROCm\textsuperscript{TM} software stack, we achieve a $10.7\times$ speedup for triangular solves and a $3.2\times$ speedup for the ILU0-preconditioned biconjugate gradient stabilized (BiCGSTAB) solver on the AMD Instinct\textsuperscript{TM} MI210 GPU.

\end{abstract}
% Copyright Statement
% When submitting your final paper to a SIAM proceedings, it is requested that you include
% the appropriate copyright in the footer of the paper.  The copyright added should be
% consistent with the copyright selected on the copyright form submitted with the paper.
% Please note that "20XX" should be changed to the year of the meeting.

% Default Copyright Statement
%\fancyfoot[R]{\scriptsize{Copyright \textcopyright\ 2025 by SIAM\\
%Unauthorized reproduction of this article is prohibited}}

% Depending on which copyright you agree to when you sign the copyright form, the copyright
% can be changed to one of the following after commenting out the default copyright statement
% above.

%\fancyfoot[R]{\scriptsize{Copyright \textcopyright\ 20XX\\
%Copyright for this paper is retained by authors}}

%\fancyfoot[R]{\scriptsize{Copyright \textcopyright\ 20XX\\
%Copyright retained by principal author's organization}}

%\pagenumbering{arabic}
%\setcounter{page}{1}%Leave this line commented out.

\section{Introduction}
\label{sec:intro}
\begin{figure}[h]
\vspace{-6pt}
    \centering
    \includegraphics[width=0.42\textwidth]{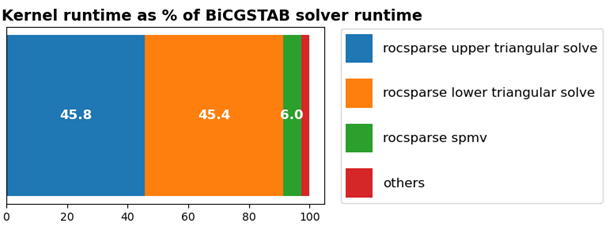} % Replace with the correct file path
    \vspace{-6pt}
	\caption{Runtime breakdown of BiCGSTAB on AMD Instinct\textsuperscript{TM} MI210. }
    \label{fig:kernel_runtime}
    \vspace{-7pt}
\end{figure}

Partial differential equations (PDEs) model complex phenomena in fields such as astrophysics~\cite{space}, finance~\cite{finance}, and fluid dynamics~\cite{spe10paper}. Discretizing PDEs using time-implicit methods generates systems of sparse linear systems, typically solved using iterative methods. To accelerate convergence, iterative solvers often rely on preconditioners applied through triangular solves, which introduce challenges due to irregular memory access and data dependencies.
Figure~\ref{fig:kernel_runtime} shows the kernel runtime breakdown for preconditioned biconjugate gradient stabilized (BiCGSTAB) solver in the Open Porous Media (OPM) simulator~\cite{opm} on the AMD Instinct\textsuperscript{TM} MI210 GPU. The solver operates on a $3\,\times\,3$ block sparse row (BSR) matrix from a 3D Laplacian~~\cite{laplacian} with 4M rows and 28M columns. Over 90\% of the runtime is spent in \texttt{rocSPARSE}~\cite{rocsparse} triangular solve kernels.
% Figure~\ref{fig:kernel_runtime} shows the kernel runtime breakdown for the GPU-based preconditioned BiConjugate Gradient Stabilized (BiCGSTAB) solver from the Open Porous Media (OPM) simulator~\cite{opm}, using a $3 \times 3$ block sparse row (BSR) matrix from a 3D Laplacian discretization with 4 million rows and 28 million columns. On the AMD MI210 GPU, over 90\% of the runtime is spent in \texttt{rocSPARSE}~\cite{rocsparse} triangular solve kernels.
%: \texttt{bsrsv\_lower\_shared} and \texttt{bsrsv\_upper\_shared}.
% Figure~\ref{fig:kernel_runtime} shows the kernel runtime breakdown for the GPU-based preconditioned BiConjugate Gradient Stabilized (BiCGSTAB) solver from the Open Porous Media (OPM) simulator~\cite{opm}, which operates on a block sparse row (BSR) matrix with \(3 \times 3\) blocks. The input used is discrtization of 3D Laplacian with 4 million rows and 28 million columns.  Over 90\% of the total runtime on the AMD MI210 GPU is spent in the \texttt{rocSPARSE}~\cite{rocsparse} triangular solve kernels: \texttt{bsrsv\_lower\_shared} and \texttt{bsrsv\_upper\_shared}.
% \vspace{-20pt}
%presented by the sparse triangular solves.
%The solver uses the rocSPARSE library~\cite{rocsparse} to compute block triangular solves and the rocBLAS library~\cite{rocblas} for other operations such as computing norm and dot product.
Commonly used strategies for improving the performance of these iterative solvers are 
%as follows:
%listed below 
% \begin{enumerate}[leftmargin=*]
%     \item Synchronization-free optimizations for parallel triangular solves~\cite{liu1,Liu2}.
%     \item DAG-based triangular solves with explicit dependency mapping~\cite{helal,Naumov}.
%     \item Solvers using reduced or mixed precision~\cite{mp1,mp2}.
%     \item Krylov subspace recycling for sequences of related systems~\cite{rbicgstab,parks_and_de_sturler}.
%     \item Domain decomposition techniques for distributed systems~\cite{ddbook,saad}.
% \end{enumerate}
(1) synchronization-free parallel algorithms~\cite{liu1,Liu2}, (2) DAG-based solves with explicit dependency mapping~\cite{helal,Naumov}, (3) reduced and mixed-precision solvers~\cite{mp1,mp2}, (4) Krylov subspace recycling~\cite{rbicgstab,parks_and_de_sturler}, and (5) domain decomposition techniques for distributed systems~\cite{ddbook,saad}; see~\S\ref{sec:related} for details on this related work.

% \begin{enumerate}[leftmargin=*]
%     \item Synchronization-free parallel triangular solve optimizations~\cite{liu1,Liu2}.
%     \item DAG-based solves with explicit dependency mapping~\cite{helal,Naumov}.
%     \item Reduced and mixed-precision solvers~\cite{mp1,mp2}.
%     \item Krylov subspace recycling~\cite{rbicgstab,parks_and_de_sturler}.
%     \item Domain decomposition for distributed systems~\cite{ddbook,saad}.
% \end{enumerate}

%This work presents a GPU-centric domain decomposition approach to accelerate triangular solves. The input matrix is partitioned into subdomains, and all inter-subdomain nonzeros are explicitly removed. This yields a sparsity pattern with no dependencies across partitions, enabling parallel triangular solves. 
%However, the resulting preconditioner is weaker, potentially increasing the number of iterations required for solver convergence.
Some combination of the above optimizations is typically used with domain decomposition, which traditionally maps each subdomain to a separate node in multi-node systems.
% Some combination of the aforementioned optimizations is typically used together with domain decomposition methods. 
% Traditional domain decomposition maps each subdomain to a separate node in multi-node systems. 
In contrast, we propose a fine-grained domain decomposition approach such that when computing triangular solves, %(1) 
we map each subdomain to a compute unit (CU) on a GPU, enabling the use of fast shared memory for vector data, 
and confine dependency resolution within CUs
%and (2) dependency resolution is confined within CUs, performed 
via GPU thread block-level synchronization or busy-waiting, avoiding costly inter-block dependency resolution.
Fine-grained domain decomposition increases parallelism in sparse triangular solves at the cost of more iterations for convergence. This work examines the effect of fine-grained decomposition on the performance and convergence of BiCGSTAB solver.
In all, we make the following contributions in this work:
%\atharva{replace SM with compute unit everywhere}
\begin{itemize}[leftmargin=*]

%\begin{itemize}
    \item GPU-centric domain decomposition 
    %strategy 
    that maps subdomains to individual compute units (CUs) while computing triangular solves.
    \item Multi-dimensional optimizations for triangular solves on matrices with subdomains, including:
    \begin{itemize}[leftmargin=*]
        \item Domain decomposition with subdomain sizes tuned to fit vector data in shared memory, reducing irregular accesses.
        \item Spin-loop and DAG-based triangular solves for matrices with non-overlapping subdomains.
        \item Fused lower/upper triangular solves within subdomains.

        % \item Domain decomposition with partition sizes that allow corresponding vector data to fit entirely in shared memory, reducing irregular global memory accesses.
        %  \item Investigating spin-loop-based and directed acyclic graph (DAG)-based approaches for triangular solves on matrices with non-overlapping subdomains.
        % \item Fusing lower and upper triangular solves within a subdomain.
       
    \end{itemize}
    
    \item Sparse triangular solves on subdomains with associated optimizations achieving:
    \begin{itemize}[leftmargin=*]
        \item \(10.7\times\) geometric mean speedup over \texttt{rocSPARSE} for triangular solve on AMD Instinct MI210 GPU.
        \item Overall \(3.2\times\) geometric mean speedup for  BiCGSTAB solver on AMD Instinct MI210 GPU, despite a \(1.6\times\) increase in iteration count.
        \item Near-linear speedup  for triangular solve kernel on up to 8 AMD Instinct MI210 GPUs for problems large enough to saturate all GPUs.
\end{itemize}
    % \begin{itemize}
    %     \item A geometric mean speedup of \(10.7\times\) over \texttt{rocSPARSE} sparse triangular solves.
    %     \item Overall \(3.2\times\) geometric mean speedup for the BiCGSTAB solver on AMD MI210 GPU, despite a \(1.6\times\) increase in iteration count.
    %     \item Near-linear speedup for our triangular solve kernels on up to 8 AMD MI 210 GPUs for large graphs
    % \end{itemize}
\end{itemize}
We evaluate the impact of fine-grained domain decomposition on both the performance and convergence of a GPU-based BiCGSTAB solver using compressed sparse row (CSR) and BSR inputs.
For BSR, we use a block size of $3\,\times\,3$, motivated by oil and gas applications where PDE discretizations yield three unknowns per cell.
To the best of our knowledge, domain decomposition at this level of granularity has \emph{not} been previously explored.

\section{Background}
\label{sec:backgroud}
In iterative solvers, a preconditioner is applied to the intermediate solution to accelerate convergence by reducing the iteration count~\cite{opm, saad}. 
This work employs the ILU0 preconditioner (i.e., incomplete LU factorization with 0 fill-in) for the BiCGSTAB solver.
ILU0-preconditioned BiCGSTAB is an important solver in many industrial applications, for example, oil and gas applications such as reservoir simulations~\cite{opm}.
%\thomas{I think this would be a good place to add a sentence emphasizing that ILU0/triangular solvers are still an industry standard in many contexts, hence why it was selected here. This is to try to avoid the "why not use X preconditioner instead". KJ I think can speak to this more accurately in the context of Oil \& Gas.}

Algorithm~\ref{alg:BiCGSTAB_with_triangular_solves} describes the implementation of a preconditioned BiCGSTAB~\cite{saad} solver.
This work uses the BiCGSTAB implementation from the Open Porous Media (OPM) initiative~\cite{opm} as the baseline and evaluates the impact of domain decomposition and associated optimizations for sparse triangular solves on the performance and convergence of the solver.

% \begin{wrapfigure}{r}{0.8\columnwidth}
% \vspace{-10pt}
% \begin{minipage}{0.8\columnwidth}
% {\small
% \begin{algorithm}
% \caption{BiCGSTAB with preconditioning~\cite{vandervost-bicgstab,saad}}
% \label{alg:BiCGSTAB_with_triangular_solves}
% \KwIn{Matrix $A$, right-hand side $b$, initial guess $x_0$, lower and upper preconditioners $\mathbf{K_1}$ and $\mathbf{K_2}$, and tolerance $\epsilon_0$}
% \KwOut{Solution vector $x$}

% Initialize $y_0 \gets \mathbf{K_2} x_0$ 
% Initialize $r_0 \gets \mathbf{K_1}^{-1}b - \mathbf{K_1}^{-1} A \mathbf{K_2}^{-1} y_0$ 
% Initialize $p_0 \gets r_0$, $\rho_0 \gets (r_0 \cdot r_0)$\;

% \For{$j \gets 0$ \KwTo \textbf{maximum iterations}}{
%     $v_j \gets \mathbf{K_1}^{-1} A \mathbf{K_2}^{-1} p_j$ \tcp*[r]{rocSPARSE triangular solves~\cite{rocsparse}}
%     $\alpha_j \gets \rho_j / (v_j \cdot r_0)$\;
%     $s_j \gets r_j - \alpha_j v_j$\;
%     $t_j \gets \mathbf{K_1}^{-1} A \mathbf{K_2}^{-1} s_j$ \hspace{-10}\tcp*[r]{rocSPARSE triangular solves~\cite{rocsparse}}
%     $\omega_j \gets (t_j \cdot s_j) / (t_j \cdot t_j)$\;
%     $y_{j+1} \gets y_j + \alpha_j p_j + \omega_j s_j$\;
%     $r_{j+1} \gets s_j - \omega_j t_j$\;

%     \If{$\|r_{j+1}\| < \epsilon_0$}{
%         \textbf{break}\;
%     }

%     $\rho_{j+1} \gets (r_{j+1} \cdot r_0)$\;
%     $\beta_j \gets (\alpha_j / \omega_j)(\rho_{j+1} / \rho_j)$\;
%     $p_{j+1} \gets r_{j+1} + \beta_j \left(p_j - \omega_j v_j\right)$\;
% }

% $x_{j+1} \gets \mathbf{K_2}^{-1} y_{j+1}$ \tcp*[r]{rocSPARSE triangular solve~\cite{rocsparse}}

% \end{algorithm}
% }
\begin{algorithm}
{\small
\caption{Preconditioned BiCGSTAB~\cite{vandervost-bicgstab,saad}}
\label{alg:BiCGSTAB_with_triangular_solves}
\KwIn{Matrix $A$, right-hand side $b$, initial guess $x_0$, lower and upper preconditioners $\mathbf{K_1}$ and $\mathbf{K_2}$, and tolerance $\epsilon_0$}
\KwOut{Solution vector $x$}

Initialize $y_0 \gets \mathbf{K_2} x_0$\\
Initialize $r_0 \gets \mathbf{K_1}^{-1}b - \mathbf{K_1}^{-1} A \mathbf{K_2}^{-1} y_0$ \\
Initialize $p_0 \gets r_0$, $\rho_0 \gets (r_0 \cdot r_0)$\;\\
\For{$j \gets 0$ \KwTo \textbf{maximum iterations}}{
    $v_j \gets \mathbf{K_1}^{-1} A \mathbf{K_2}^{-1} p_j$ \tcp*[l]{rocSPARSE triangular solves~\cite{rocsparse}}
    $\alpha_j \gets \rho_j / (v_j \cdot r_0)$\\
    $s_j \gets r_j - \alpha_j v_j$\\
    $t_j \gets \mathbf{K_1}^{-1} A \mathbf{K_2}^{-1} s_j$ \hspace{-10pt}\tcp*[r]{rocSPARSE triangular solves~\cite{rocsparse}}
    $\omega_j \gets (t_j \cdot s_j) / (t_j \cdot t_j)$\\
    $y_{j+1} \gets y_j + \alpha_j p_j + \omega_j s_j$\\
    $r_{j+1} \gets s_j - \omega_j t_j$\;

    \If{$\|r_{j+1}\| < \epsilon_0$}{
        \textbf{break}\;
    }

    $\rho_{j+1} \gets (r_{j+1} \cdot r_0)$\\
    $\beta_j \gets (\alpha_j / \omega_j)(\rho_{j+1} / \rho_j)$\\
    $p_{j+1} \gets r_{j+1} + \beta_j \left(p_j - \omega_j v_j\right)$\;
}

$x_{j+1} \gets \mathbf{K_2}^{-1} y_{j+1}$ \tcp*[r]{rocSPARSE triangular solve~\cite{rocsparse}}
}
\end{algorithm}
% \end{minipage}
% \vspace{-10pt}
% \end{wrapfigure}

We profile read/write bandwidths of the most expensive kernels in the ILU0-preconditioned BiCGSTAB solver~\cite{opm} on an AMD Instinct\textsuperscript{TM} MI210 GPU using the \texttt{rocprof}~\cite{rocprof} profiler. As shown in Table~\ref{tab:bwintro}, triangular solve kernels use only a small fraction of peak bandwidth, while kernels such as \texttt{rocblas\_dot} reach up to 75\%.
This evaluation highlights the limitations of triangular solve kernels in effectively utilizing the available memory bandwidth.
We address these bottlenecks by using domain decomposition. The next section, \S\ref{sec:dd}, presents our approach for domain decomposition.

%While we use BiCGSTAB + ILU0 as a case study, our approach applies to any preconditioner (ex. Cholesky, ILU-K) and  preconditioned iterative solver (ex. Gauss Seidel, Conjugate gradient) relying on triangular solves.
%\atharva{push this to intro. Mention that this is also applicable when used as a smoother}
%and algebraic multigrid (AMG) solvers.
%iterative solver with incomplete cholesky 
\begin{table}[!ht]
\centering
\caption{Read/Write Bandwidth of Five Most Expensive Kernels in OPM BiCGSTAB on the AMD Instinct\textsuperscript{TM} MI210 GPU~\cite{opm}.
The input matrix is a discretized Laplacian with 4-million rows \& 28-million non-zeros. The peak read/write theoretical bandwidths on
AMD Instinct\textsuperscript{TM} MI210 GPU is 1.6 TB/s~\cite{mi210_bandwidth}. The kernel \texttt{bsrxmv\_3x3\_kernel} refers to the
block sparse matrix-vector multiply computation. }
\label{tab:kernel_bandwidth}
\resizebox{\columnwidth}{!}
{
\begin{tabular}{|l|l|l|}
\hline
Kernel Name                & \begin{tabular}[c]{@{}l@{}}Read \\ Bandwidth \\ (GB/s)\end{tabular} & \begin{tabular}[c]{@{}l@{}}Write \\ Bandwidth \\ (GB/s)\end{tabular} \\ \hline
bsrsv\_upper\_shared       & 229.23                                                              & 23.82                                                                \\ \hline
bsrsv\_lower\_shared       & 230.88                                                               & 24.14                                                                \\ \hline
bsrxmv\_3x3\_kernel        & 832.55                                                              & 31.02                                                                \\ \hline
rocblas\_axpy\_kernel      & 816.71                                                              & 395.54                                                               \\ \hline
rocblas\_dot\_kernel\_inc1 & 1291.03                                                             & 2.83                                                                 \\ \hline
\end{tabular}
}

%  \begin{threeparttable}
%  \centering
% \begin{tablenotes}
% \footnotesize
% % \vspace{-3pt}
%  \item \footnotesize{Input: Discretized Laplacian with 4-million rows \& 28-million non-zeros.}
%  \item The peak read bandwidth on AMD Instinct\textsuperscript{TM} MI210 GPU is 1.6 TB/s~\cite{mi210_bandwidth}. 
%  \item \texttt{bsrxmv\_3x3\_kernel} refers to the block sparse matrix-vector multiply. 
% \end{tablenotes}
% \end{threeparttable}
\label{tab:bwintro}
\end{table}

aOur work differs from the existing work on sparse triangular solves and domain decomposition (as detailed in \S\ref{sec:related}) in the following ways:
\setlength{\itemsep}{0pt}
\begin{itemize}[leftmargin=*]

%\begin{enumerate}[wide, labelwidth=!, labelindent=0pt]
%\begin{enumerate}
    \item \textbf{GPU-oriented domain decomposition:} While domain decomposition has been explored previously, we focus on its effects with a GPU architecture in mind. We design subdomains such that each subdomain is mapped to a GPU compute unit.
    %, with each thread block processing a unique subdomain. 
   % and improving preconditioner application efficiency. %Additionally, we examine how DAG-based and synchronization-free methods perform for triangular olves on matrices with independent subdomains.
    This approach increases parallelism but requires additional solver iterations.
    \item \textbf{Optimizations for triangular solves:} Prior work on optimizing synchronization-free and DAG-based methods respect the dependencies between rows to correctly apply preconditioning. 
    In contrast, 
    %our approach for preconditioner design divides the matrix into multiple partitions (subdomains) and deliberately drops the connections (nonzeros) between partitions. 
    we explore optimizations for triangular solves using synchronization-free and DAG-based methods on matrices with non-overlapping subdomains.
    To optimize 
    %for 
    performance, we keep subdomain sizes small enough for vector data to fit 
    %entirely 
    in shared memory, thus reducing irregular memory accesses to global memory.
    %This trade-off enables parallelism by confining dependencies within each subdomain, while intentionally ignoring inter-partition dependencies.. 
    
    \item \textbf{Case study with BiCGSTAB and ILU0:} 
    %In this work, 
    We use the ILU0-preconditioned BiCGSTAB as a case study to evaluate the impact of our sparse triangular-solve optimizations enabled by fine-grained domain decomposition. While this solver-preconditioner pair is used for demonstration, our approach generalizes to other iterative solvers that rely on preconditioners involving triangular solves, such as conjugate gradient and Gauss-Seidel with preconditioners, like ILU-k and incomplete Cholesky. 
    It can also extend to triangular solves used as smoothers in algebraic multigrid (AMG) methods.

%\end{enumerate}
\end{itemize}

\section{Fine-grained Domain Decomposition}
\label{sec:dd}
\setlength{\columnsep}{5pt}      % Horizontal space between wrapfig and text (default is ~10pt)

Domain decomposition methods are 
based on the principle of divide-and-conquer and are commonly used for solving partial differential equations in two- or three-dimensional domains~\cite{saad}. 
These techniques allow for added parallelism by dividing the computational domain into non-overlapping regions that can be processed simultaneously. 
However, this added parallelism often comes at the cost of additional iterations until the solver converges~\cite{anzt}. 
% \begin{wrapfigure}{r}{0.55\columnwidth}
% \vspace{-10pt}
% \begin{minipage}{0.54\columnwidth}

% \end{minipage}
% \vspace{-10pt}
% \end{wrapfigure}
The domain decomposition method employed in this work begins with partitioning the input matrix. %, a process that can be achieved through various approaches. 
When the geometry of the grid is known, simple geometric cuts along the domain can identify partition labels for matrix rows. 
Alternatively, graph-based partitioning can assign labels to vertices. 
This work leverages both techniques: (1) geometric cuts to partition the domain and reorder the input matrix and (2) graph-based partitioning using METIS~\cite{metis, metisdocs}, followed by graph reordering.

Algorithm~\ref{alg:geometric_cuts_part} presents our implementation of partitioning based on geometric cuts.
\begin{algorithm}[htbp]
\SetAlgoLined
{\small
\caption{Partitioning Using Geometric Cuts for Cartesian Grids}
\label{alg:geometric_cuts_part}
\KwIn{Grid dimensions $(n_x,n_y,n_z)$, subdomain dimensions $(nblk_x,nblk_y,nblk_z)$}
\KwOut{Partition labels \texttt{part\_id} for all vertices}
$b_x \leftarrow n_x / nblk_x$\\
$b_y \leftarrow n_y / nblk_y$\\
\For{$i \leftarrow 0$ \KwTo $n_x - 1$}{
    $ibx \leftarrow \lfloor i / nblk_x\rfloor$\\
    \For{$j \leftarrow 0$ \KwTo $n_y - 1$}{
        $jby \leftarrow \lfloor j / nblk_y\rfloor$\\
        \For{$k \leftarrow 0$ \KwTo $n_z - 1$}{
            $kbz \leftarrow \lfloor k / nblk_z\rfloor$\\
            $gidx \leftarrow i + n_x\,(j + n_y\,k)$\\
            $pidx \leftarrow ibx + b_x\,(jby + b_y\,kbz)$\\
            \texttt{part\_id}[gidx] $\leftarrow pidx$\\
        }
    }
}
}
\end{algorithm}
 
%This approach relies solely on the geometric properties of the grid, without considering the connectivity or structure of the sparse matrix. 
The algorithm maps each grid point to a unique subdomain, assigning the same label to all points within a single geometric cut.
In this work, we use a domain decomposition approach that constructs a preconditioner with non-overlapping subdomains, enabling triangular solves to be computed in parallel across the subdomains.
In addition, we use partitions of uniform size, where each partition contains the same number of rows.

\subsection{Matrix reordering}

After creating partition labels for the entire grid, we map them to the vertices of the sparse matrix representation of the grid. 
Next, we generate a row permutation by grouping matrix rows associated with unique partition labels. 
We renumber the vertices based on their labels, producing a row permutation that rearranges the rows of the matrix according to their partition labels.
\begin{algorithm}[htbp]
\SetAlgoLined
{\small
\caption{CSR Matrix Reordering Based on Row Permutation}
\label{alg:csr_matrix_reorder}
\KwIn{Original CSR matrix $A$, row permutation \texttt{pmap}[], inverse \texttt{inv\_pmap}[]}
\KwOut{Reordered CSR matrix $P$}
$nrows \leftarrow A.\texttt{nrows}$\\
$nnz \leftarrow A.\texttt{nnz}$\\
Allocate CSR matrix $P$ with $nrows$ rows and $nnz$ nonzeros\\
Initialize arrays \texttt{rowptr}, \texttt{colidx}, \texttt{values}\\
$\texttt{rowptr}[0] \leftarrow 0$\\
\For{$i \leftarrow 0$ \KwTo $nrows-1$}{
  $k \leftarrow \texttt{pmap}[i]$\\
  $\texttt{rowptr}[i+1] \leftarrow A.\texttt{rowptr}[k+1] - A.\texttt{rowptr}[k]$\\
}
\For{$i \leftarrow 0$ \KwTo $nrows-1$}{
  $\texttt{rowptr}[i+1] \leftarrow \texttt{rowptr}[i+1] + \texttt{rowptr}[i]$\\
}
$n \leftarrow 0$\\
\For{$i \leftarrow 0$ \KwTo $nrows-1$}{
  $m \leftarrow \texttt{pmap}[i]$\\
  \For{$j \leftarrow A.\texttt{rowptr}[m]$ \KwTo $A.\texttt{rowptr}[m+1]-1$}{
    $k \leftarrow \texttt{inv\_pmap}[A.\texttt{colidx}[j]]$\\
    $\texttt{colidx}[n] \leftarrow k$\\
    $\texttt{values}[n] \leftarrow A.\texttt{values}[j]$\\
    $n \leftarrow n + 1$\\
  }
  Sort \texttt{colidx}[\texttt{rowptr}[i]\dots\texttt{rowptr}[i+1]]\\
}
}
\end{algorithm}

We then apply the row permutation to the original sparse matrix, resulting in a reordered matrix where rows belonging to the same partition are arranged consecutively, ensuring that subdomains are clearly delineated. The final step involves removing the connections between the partitions. 

Algorithm~\ref{alg:csr_matrix_reorder} shows our CSR matrix reordering approach. 
For a given row $r$ in the original matrix, \texttt{pmap[$r$]} gives the reordered row index. Conversely, for a row $j$ in the reordered matrix, \texttt{inv\_pmap[$j$]} gives the corresponding row index in the original matrix.
The permutation map \texttt{pmap} is used to reorder the rows and populate the new CSR row pointers. When populating column indices, we use the inverse permutation map \texttt{inv\_pmap} to map each column index in the original matrix to its new position in the reordered matrix. %This ensures consistency between the row and column ordering after reordering.

\subsection{Removing inter-partition dependencies}
After applying the row permutation to the original matrix, we iterate over the reordered matrix and remove all nonzero elements outside the $(P \times P)$ window along the diagonal, where $P$ represents the number of rows in a partition.
We perform ILU0 factorization on the matrix generated after dropping the inter-subdomain connections (nonzeros). 
The resulting matrix from the ILU0 decomposition serves as our preconditioner, and triangular solves are applied using this preconditioner as the input matrix.
For all operations other than triangular solves, we use the partitioned and reordered matrix without dropping the inter-subdomain connections.

% \begin{algorithm}
% \caption{Partitioning using geometric cuts }
% \label{alg:geometric_cuts_part}
% \begin{algorithmic}[1]
% \STATE \textbf{Input:} Grid dimensions $(n_x, n_y, n_z)$, number of blocks $(nblk_x, nblk_y, nblk_z)$
% \STATE \textbf{Output:} Partition labels \texttt{part\_id} for all vertices
% \STATE $b_x \gets n_x / nblk_x$
% \STATE $b_y \gets n_y / nblk_y$
% \FOR{$i = 0$ \textbf{to} $n_x - 1$}
%     \STATE $ibx \gets \lfloor i / nblk_x \rfloor$
%     \FOR{$j = 0$ \textbf{to} $n_y - 1$}
%         \STATE $jby \gets \lfloor j / nblk_y \rfloor$
%         \FOR{$k = 0$ \textbf{to} $n_z - 1$}
%             \STATE $kbz \gets \lfloor k / nblk_z \rfloor$
%             \STATE $gidx \gets i + n_x \cdot (j + n_y \cdot k)$
%             \STATE $pidx \gets ibx + b_x \cdot (jby + b_y \cdot kbz)$
%             \STATE $\texttt{part\_id}[gidx] \gets pidx$
%         \ENDFOR
%     \ENDFOR
% \ENDFOR
% \end{algorithmic}
% \end{algorithm}

\begin{figure}[htb]
    \centering
    \begin{subfigure}[b]{0.15\textwidth}
        \centering
        \includegraphics[width=\textwidth]{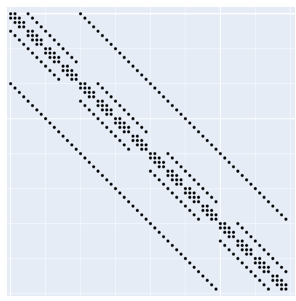}
        \caption{Discretization of 3D Laplacian}
        \label{fig:3d1}
    \end{subfigure}
    \begin{subfigure}[b]{0.15\textwidth}
        \centering
        \includegraphics[width=\textwidth]{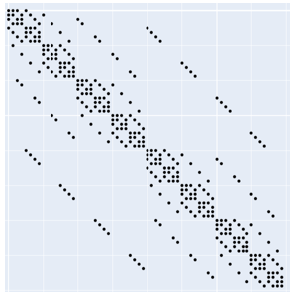}
        \caption{Partitioning \& reordering}
        \label{fig:part}
    \end{subfigure}
    \begin{subfigure}[b]{0.15\textwidth}
        \centering
        \includegraphics[width=\textwidth]{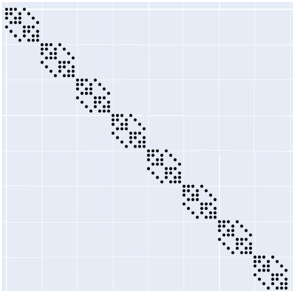}
        \caption{Without inter-partition nonzeros}
        \label{fig:ddpost}
    \end{subfigure}
    \caption{Process of domain decomposition.}
 \label{fig:processdd}
\end{figure}

Figure~\ref{fig:processdd} below illustrates the domain decomposition approach used in this work, starting from the discretization of the Laplacian equation. 
The unmodified discretization is shown in Figure~\ref{fig:3d1}. 
The input matrix is partitioned into eight partitions after computing the row permutation and reordering. The reordered matrix is shown in Figure~\ref{fig:part} above. 
Finally, nonzero elements outside the partitions are removed from the sparse matrix, and the resulting matrix used in preconditioning is shown above in Figure~\ref{fig:ddpost}.

Figure~\ref{fig:processdd} above shows a significant reduction in nonzero elements after domain decomposition. 
%as a large number of partitions results in only a small fraction of nonzeros being retained after reordering.
In our implementation, we select the size of the subdomains such that the number of nonzeros removed remains a small fraction of the total nonzeros in the original matrix.

%\section{Mapping of Subdomains to GPU Architecture}
\section{GPU-Centric Triangular Solves on Subdomains}
\label{sec:opts}

Parallel triangular solves are broadly classified into \emph{synchronization-free} and \emph{DAG-based} approaches. Synchronization-free methods rely on busy-wait loops, where threads stall until dependencies are resolved. DAG-based methods construct a dependency graph during an analysis phase and traverse it in a solution phase.
%, typically using edge- or vertex-centric strategies. 
While DAG methods expose more parallelism, they introduce nontrivial overhead during DAG construction~\cite{Naumov, liu1}.
%\thomas{Can we provide any references here?}
%\atharva{added}

Operating on matrices with non-overlapping subdomains enables optimizations that are not feasible for triangular solve library implementations designed for general matrices. 
This section details how we map the problem of triangular solves on subdomains to GPUs for both synchronization-free and DAG-based approaches.
Each subdomain is assigned to a unique thread block on the GPU.
The size of the subdomain is chosen to be small enough so that the vector data associated with it can fit entirely in shared memory.
%     The size of the subdomain is determined by the shared memory available to each thread block. 
For example, with 64~KB of shared memory per thread block on AMD Instinct\textsuperscript{TM} MI210 GPUs, the maximum subdomain size that allows storing vector data entirely in shared memory is 8,192 rows of double-precision data.
This section also explores additional optimization opportunities via ILDU0 decomposition and kernel fusion.

% We extend Algorithms~\ref{alg:lower_triangular_solve} and~\ref{alg:parallel_triangular_solve} as follows:
% \begin{enumerate}
%     \item Each subdomain is assigned to a unique thread block on the GPU.
%     \item The size of the subdomain is chosen to be small enough so that the vector data associated with it can fit entirely in shared memory.
%     The size of the subdomain is determined by the shared memory available to each thread block. 
%     For example, with 64~KB of shared memory per thread block on AMD GPUs, the maximum subdomain size that allows storing vector data entirely in shared memory is 8,192 rows for double-precision data.
% \end{enumerate}

\subsection{Synchronization-free triangular solves on non-overlapping subdomains}
% This work uses synchronization-free sparse triangular solves from \texttt{rocSPARSE}~\cite{rocsparse} library as the baseline.

% Algorithm~\ref{alg:lower_triangular_solve} outlines the implementation of a lower triangular solve on a dense matrix using a synchronization-free approach. 
% In this approach, each thread is assigned to a row of the matrix and enters a busy-wait state until the dependencies on previous rows are resolved. Figure~\ref{fig:spin1} illustrates the assignment of parallel threads to rows in the sparse matrix.
% Threads within a wavefront iterate over the nonzero elements of the row and update the result as soon as the required dependencies are resolved. 

% Triangular solves from \texttt{rocSPARSE}~\cite{rocsparse} are used as the baseline kernels in this work. 
% The baseline approach assigns wavefronts to rows of the matrix for parallel execution. Figure~\ref{fig:spin2} demonstrates the assignment of wavefronts to the rows of the matrix.
%  %\vspace{-10}
% We extend Algorithm~\ref{alg:lower_triangular_solve} to implement synchronization-free triangular solves, where each subdomain is assigned to a thread block. 
% The vector data corresponding to each subdomain is loaded into shared memory.
This work uses the synchronization-free sparse triangular solve kernels from the \texttt{rocSPARSE}~\cite{rocsparse} library as the baseline for comparison.
Figure~\ref{fig:spin1} below shows the design of a 
%generic 
parallel triangular solve, where each thread is assigned to a row of the lower triangular matrix and enters a spin loop, waiting for dependent rows to complete. Figure~\ref{fig:spin2} below shows the wavefront-based variant used in \texttt{rocSPARSE}, where wavefronts (group of threads) are assigned to rows and iterate over nonzero elements once dependencies are resolved.

\begin{figure}[H]
    \centering
    \begin{subfigure}[b]{0.22\textwidth}
        \centering
        \includegraphics[width=\textwidth]{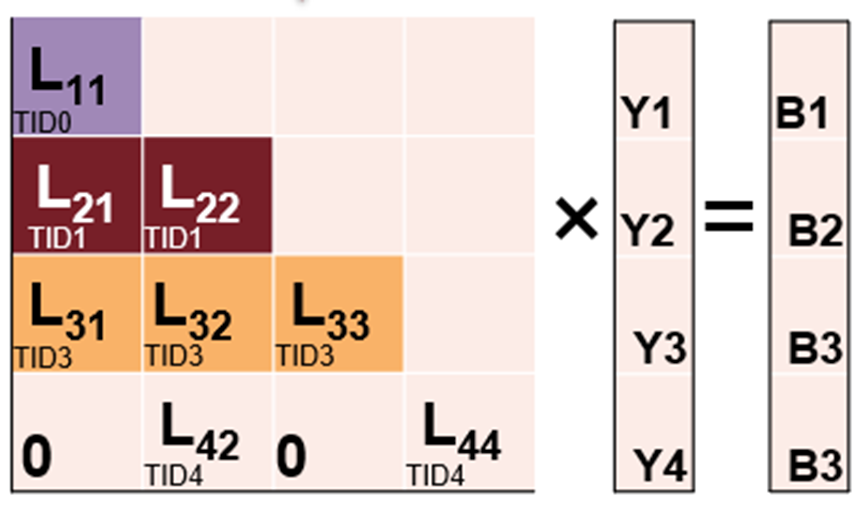}
        \caption{Lower triangular solve with threads assigned to rows.}
        \label{fig:spin1}
    \end{subfigure}
    \hfill
    \begin{subfigure}[b]{0.23\textwidth}
        \centering
        \includegraphics[width=\textwidth]{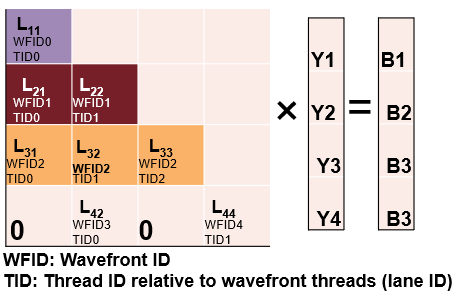}
        \caption{Triangular solve with wavefronts assigned to rows.}
        \label{fig:spin2}
    \end{subfigure}
    \caption{Lower triangular solves: threads/wavefronts assigned to rows.}
    \label{fig:triangular_solve_strategies}
\vspace{-4pt}
\end{figure}

We extend this baseline in two key ways:
(1) each subdomain is explicitly assigned to a thread block, and
(2) partial sums are computed in shared memory. This is feasible because subdomain sizes are selected such that a shared array of size equal to the number of subdomain rows can fit entirely in shared memory (local data share or LDS —-- the register-adjacent memory on AMD GPUs).

Algorithm~\ref{alg:lower_triangular_solve} below outlines our implementation of synchronization-free triangular solves using spin loops and shared memory. In our approach, the algorithm is applied to sparse matrices stored in either CSR or BSR formats. At runtime, each thread block is mapped to a subdomain, each wavefront in a thread block is assigned to a row, and threads within a wavefront process the row's nonzero columns in parallel. For BSR inputs with $3\,\times\,3$ blocks, each nonzero block, consisting of nine elements, is loaded into memory that is local to the thread responsible for processing it. This allows threads to have fast accesses to all the elements of the block during computation, minimizing latency from memory accesses.
% per thread to improve memory locality and reuse.
\begin{algorithm}[H]
\small
\caption{Lower Triangular Solve with Spin Loops and Shared Partial Sums}
\label{alg:lower_triangular_solve}
\KwIn{Lower triangular matrix $L$, RHS vector $b$, subdomain partitioning, per‐subdomain \texttt{done} flags (initialized to 0)}
\KwOut{Solution vector $x$ (initialized to zero before call)}

\tcp{Process each subdomain in parallel}
\ForEach{subdomain $s$ in parallel}{
  \tcp{Per‐row partial sums buffer}
  \textbf{shared:} \texttt{sum}[]

  \tcp{Zero out per‐row partial sums}
  \ForEach{row $i\in s$ in parallel}{
    \texttt{sum}[$\,i - s_{\mathrm{start}}$] $\gets 0$\;
  }
  \_\_syncthreads()\;

  \tcp{Compute each row via spin‐loops}
  \ForEach{row $i\in s$ in parallel}{
    \For{each $j<i$ with $L[i][j]\neq 0$}{
      \While{\textbf{not}\;atomic\_load(\texttt{done}[j])}{
        \textbf{continue}\;
      }
      \texttt{sum}[$\,i - s_{\mathrm{start}}$] $\;+\!=\;L[i][j]\times x[j]$\;
    }
    \tcp{Update row solution and done array }
    $x[i]\gets(b[i]-\texttt{sum}[\,i - s_{\mathrm{start}}\,])\,/\,L[i][i]$\;
    atomic\_store(\texttt{done}[i],1)\;
  }
}
\end{algorithm}

%The same logic is extended to sparse matrices. For sparse matrices, the algorithm for GPU modifies the assignment such that each wavefront (a group of 64 threads on AMD GPUs) is assigned to a row. 
%Additionally, we explore the performance impact of reducing the subdomain size further, allowing both the vector data and the dependency tracking array (\texttt{done array} in Algorithm~\ref{alg:lower_triangular_solve}) to be stored entirely in shared memory.
While spin-loop-based variants avoid the need for explicit thread block-wide or GPU-wide synchronization, they suffer from excessive active waiting. Threads assigned to rows with deep dependency chains may remain in a spin loop for extended periods, leading to underutilized compute resources and  memory bandwidth~\cite{helal}.

As an alternative to the synchronization-free approach, we also explore DAG-based triangular solves that use explicit dependency tracking to coordinate computation across levels.

\subsection{DAG-based triangular solves on subdomains}
\begin{figure}[hb]
    \centering
    \begin{subfigure}[b]{0.25\textwidth}
        \centering
        \includegraphics[width=\textwidth]{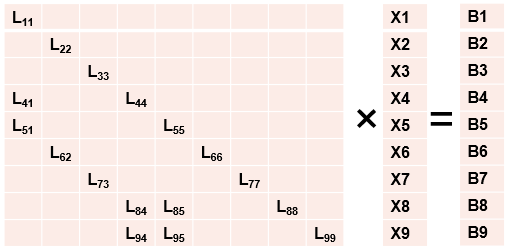}
        \caption{System of sparse linear equations}
        \label{fig:d1}
    \end{subfigure}
    \hfill
    \begin{subfigure}[b]{0.2\textwidth}
        \centering
        \includegraphics[width=\textwidth]{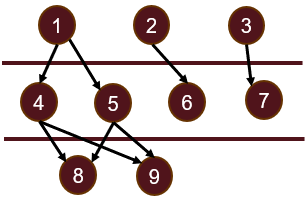}
        \caption{DAG based on the dependencies}
        \label{fig:d2}
    \end{subfigure}
    \caption{DAG construction for a sparse linear system.}
    \label{fig:dagts}
\end{figure}

Level-set or DAG-based approaches explicitly track dependencies in the form of a directed acyclic graph (DAG). Figure~\ref{fig:dagts} on the next page
illustrates the construction of a DAG corresponding to a system of sparse linear equations. 
Level 0 of the DAG contains all rows that can be updated at the start of the triangular solve, as these rows do not have off-diagonal nonzero entries. 
Subsequent levels are formed by adding rows that depend only on rows from earlier levels. To minimize the overhead associated with level assignment, we take advantage of 
the fact that there are no dependencies between subdomains 
\begin{algorithm}[H]
{\small
\caption{GPU-based parallel level assignment for lower triangular matrix for subdomain rows }
\label{alg:level_assignment}
\KwIn{Lower triangular matrix $L$}
\KwOut{level array \texttt{hmap}, initialized with all elements = $\#$rows +1}

$tid \gets \texttt{threadIdx.x}$,\quad $bdim \gets \texttt{blockDim.x}$\;
$start, end \gets$ subdomain row range\;
\textbf{shared:} $\texttt{marked}[\,]$, $\texttt{level} = 0$, $\texttt{added} = 0$\;

\tcp{Initialize: mark level 0 for diagonal-only rows}
\For{$i \gets start + tid,\ i < end;\ i += bdim$}{
  $\texttt{marked}[i - start] \gets 0$\;\\
  \If{$L[i][j] = 0$ for all $j < i$}{
    $\texttt{hmap}[i] \gets 0$\;
  }
}
\texttt{\_\_syncthreads()}\;

\While{true}{
  \If{$tid == 0$}{ $\texttt{added} \gets 0$ }
  \texttt{\_\_syncthreads()}\;

  \tcp{Check eligibility of rows for current level}
  \For{$i \gets start + tid,\ i < end;\ i += bdim$}{
    \If{$\texttt{hmap}[i] > \texttt{level}$}{
      $valid \gets \texttt{true}$\;
      \For{$j < i$ such that $L[i][j] \ne 0$}{
        \If{$\texttt{hmap}[j] > \texttt{level}$}{
          $valid \gets \texttt{false}$; \textbf{break}
        }
      }
      \If{$valid$}{
        $\texttt{marked}[i - start] \gets 1$\;
        \texttt{atomicOr(\&added, 1)}\;
      }
    }
  }
  \texttt{\_\_syncthreads()}\;

  \tcp{Promote eligible rows to next level}
  \For{$i \gets start + tid,\ i < end;\ i += bdim$}{
    \If{$\texttt{marked}[i - start]$}{
      $\texttt{hmap}[i] \gets \texttt{level} + 1$, $\texttt{marked}[i - start] \gets 0$\;
    }
  }
  \texttt{\_\_syncthreads()}\;

  \If{$added == 0$}{ \textbf{break} }
  \If{$tid == 0$}{ $\texttt{level}++$ }
  \texttt{\_\_syncthreads()}
}
}
\end{algorithm}

and perform level assignments for each subdomain in parallel on the GPU. 
This approach significantly reduces the overhead associated with sequential level assignment.

\subsubsection{Parallel level assignment for subdomain rows} Algorithm~\ref{alg:level_assignment} shown on the previous page outlines
the construction of the level set of the DAG on GPU. 
Within a subdomain, which is assigned to a GPU thread block, we identify rows that contain only diagonal nonzeros and assign them a level of 0. The main loop then iteratively searches for rows whose dependencies have already been assigned a level. A shared array, \texttt{marked}, is used to track rows that are eligible for promotion to the next level during each iteration.

After marking all eligible rows, we perform a block-wide synchronization to prevent race conditions before updating the level assignments. The level map is then incremented for all marked rows. To determine whether progress has been made in the current iteration, we use an \texttt{atomicOr} operation on a shared flag, which signals if any row was marked for update. The algorithm terminates when no new rows are marked for level assignment.
We perform this process for all subdomains in parallel.

\subsubsection{Parallelization of DAG-based solves on subdomains}
Algorithm~\ref{alg:parallel_triangular_solve} describes the parallelization of triangular solves using a DAG. 
All rows within a current level are updated in parallel, and synchronization between levels is required to ensure correctness. This work evaluates the performance of DAG-based triangular solves on matrices with  non-overlapping subdomains.

Instead of using a conditional statement to check which rows need to be processed at the current level, we explicitly construct a version of the directed acyclic graph (DAG) that tracks both the number of rows to be processed at a given level and the corresponding row indices.

%\vspace{-10}
% \begin{algorithm}
% {\small
% \caption{DAG-based lower triangular solve on GPU using shared memory}
% \label{alg:parallel_triangular_solve}
% \KwIn{Lower triangular matrix $L$, input vector $b$, level array \texttt{hmap}, subdomain partitioning}
% \KwOut{Solution vector $x$}
% \tcp{Subdomain assigned to   thread block}
% \ForEach{subdomain $s$ \textbf{in parallel}}{
%     \textbf{shared:} $\texttt{sum}[\,]$ \tcp*{Buffer for forward substitution}

%     \ForEach{row $i \in s$ \textbf{in parallel}}{
%         $\texttt{sum}[i - s_{\text{start}}] \gets b[i]$\;
%     }
%     \texttt{\_\_syncthreads()}

%     $max\_level \gets \max(\texttt{hmap}[i])$ for $i \in s$\;

%     \For{$\ell \gets 0$ \KwTo $max\_level$}{
%         \ForEach{row $i \in s$ \textbf{such that} $\texttt{hmap}[i] = \ell$ \textbf{in parallel}}{
%             \For{$j < i$ such that $L[i][j] \ne 0$}{
%                 $\texttt{sum}[i - s_{\text{start}}] \mathrel{-}= L[i][j] \cdot \texttt{sum}[j - s_{\text{start}}]$\;
%             }
%             $\texttt{sum}[i - s_{\text{start}}] \mathrel{/}= L[i][i]$\;
%         }
%         \texttt{\_\_syncthreads()}
%     }

%     \tcp{Write final result to global memory}
%     \ForEach{row $i \in s$ \textbf{in parallel}}{
%         $x[i] \gets \texttt{sum}[i - s_{\text{start}}]$\;
%     }
% }
% \end{algorithm}

% }
\begin{algorithm}
{\small
\caption{DAG-based lower triangular solve on GPU using shared memory}
\label{alg:parallel_triangular_solve}
\KwIn{Lower triangular matrix $L$, input vector $b$, level array \texttt{hmap}, subdomain partitioning}
\KwOut{Solution vector $x$}
\tcp{Subdomain assigned to   thread block}
\ForEach{subdomain $s$ \textbf{in parallel}}{
    \textbf{shared:} $\texttt{sum}[\,]$ \tcp*{Buffer for forward substitution}

    \ForEach{row $i \in s$ \textbf{in parallel}}{
        $\texttt{sum}[i - s_{\text{start}}] \gets b[i]$\;
    }
    \texttt{\_\_syncthreads()}

    $max\_level \gets \max(\texttt{hmap}[i])$ for $i \in s$\;

    \For{$\ell \gets 0$ \KwTo $max\_level$}{
        \ForEach{row $i \in s$ \textbf{such that} $\texttt{hmap}[i] = \ell$ \textbf{in parallel}}{
            \For{$j < i$ such that $L[i][j] \ne 0$}{
                $\texttt{sum}[i - s_{\text{start}}] \mathrel{-}= L[i][j] \cdot \texttt{sum}[j - s_{\text{start}}]$\;
            }
            $\texttt{sum}[i - s_{\text{start}}] \mathrel{/}= L[i][i]$\;
        }
        \texttt{\_\_syncthreads()}
    }

    \tcp{Write final result to global memory}
    \ForEach{row $i \in s$ \textbf{in parallel}}{
        $x[i] \gets \texttt{sum}[i - s_{\text{start}}]$\;
    }
}
}
\end{algorithm}

% In our implementation using domain decomposition, we extend Algorithms~\ref{alg:level_assignment} and \ref{alg:parallel_triangular_solve} in the following ways:

% \begin{enumerate}[leftmargin=*]
%     \item We take advantage of the fact that there are no dependencies between subdomains and perform level assignment for each subdomain in parallel on the GPU. 
%     This approach significantly reduces the overhead associated with sequential level assignment.
%     \item Instead of using a conditional statement to check which rows need to be processed at the current level, we explicitly construct a version of the directed acyclic graph (DAG) that tracks both the number of rows to be processed at a given level and the corresponding row indices.
% \end{enumerate}

%\atharva{Algorithms 5-7: I managed to parallelize DAG generation on the GPU. TODO: Update the algorithms 5-7 to show execution on subdomains.  }
%\vspace*{-6pt}
Figure~\ref{fig:dagblocks} below shows the generation of DAGs for matrices with subdomains. Instead of constructing a DAG for the entire matrix, we create separate DAGs for each subdomain. 
Since there are no dependencies outside of the subdomains, the DAG for each subdomain is disconnected from the DAGs of other subdomains. 
We assign thread blocks to the DAGs corresponding to each subdomain. Synchronization between successive levels of the DAG is achieved using thread block-wide synchronization via  
(\texttt{\_\_syncthreads}). As a result, DAG-based triangular solves on matrices with subdomains can be performed within a single kernel invocation from the host CPU.
We explore the impact of two variants of DAG traversal:
\subsubsection{Vertex-Centric DAG Traversal} 
\label{sssec:vc-dag}
After creating a DAG, threads within a thread block are assigned to all the rows in a given level. A thread operating on a given row processes the nonzeros of the row
\begin{figure}[H]
    \vspace{-6pt}
    \centering
    \includegraphics[width=0.45\textwidth]{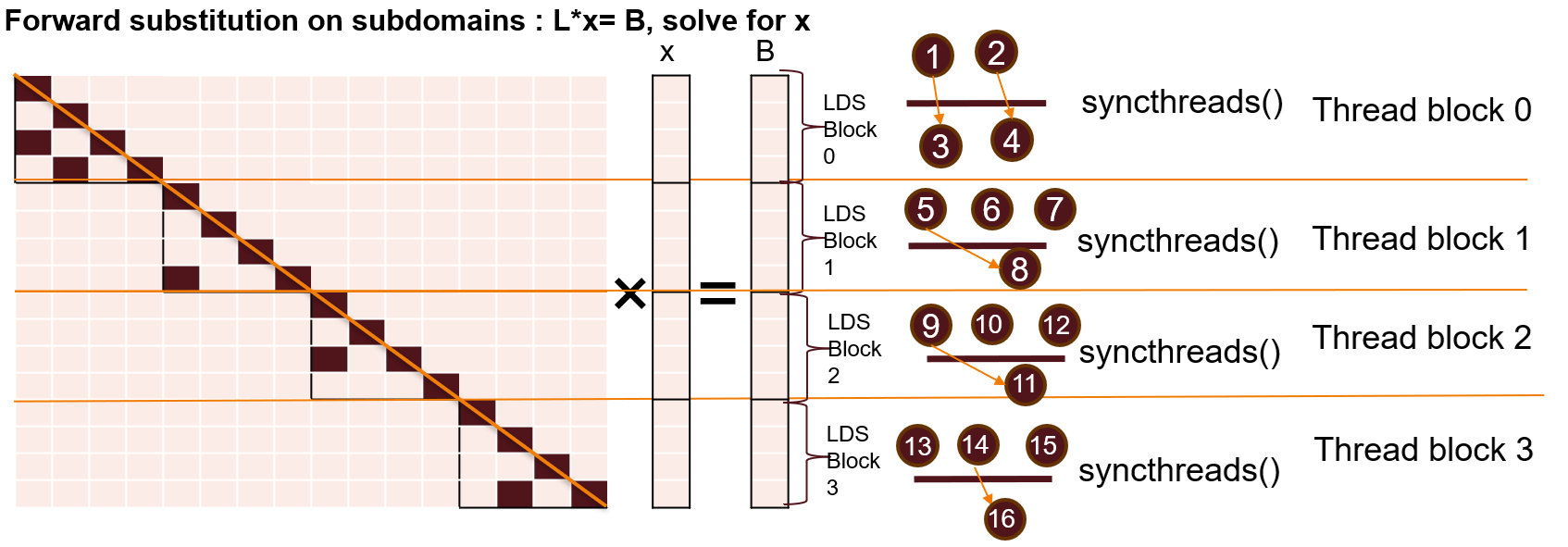}
    \vspace{-6pt}
    \caption{DAG-based triangular solves on subdomains.}
    \label{fig:dagblocks}
\end{figure}
\noindent serially. This approach is conceptually similar to vertex-centric parallelism, as explored in graph workloads, where the neighbors of a vertex are processed sequentially by each thread.

\subsubsection{Edge-Centric DAG Traversal}
\label{sssec:ec-dag}
In this approach, threads are assigned to the nonzero elements of the rows at a given level. All nonzeros are updated in parallel, requiring atomic operations to correctly compute the unknown corresponding to each row. Edge-centric parallelism, as previously explored in graph workloads, provides greater parallelism than vertex-centric approaches and achieves higher bandwidth utilization~\cite{edge_v_node}.

This work evaluates the impact of edge-centric and vertex-centric DAG traversal on both performance and convergence. Edge-centric kernels use atomic updates in a non-deterministic order, which can lead to variation in iteration counts, unlike vertex-centric kernels that compute row-wise in a fixed order. DAG-based solves can achieve higher bandwidth utilization than synchronization-free approaches~\cite{helal}, even without non-overlapping subdomains. By restricting DAG traversal to subdomains and using shared memory for vector accesses, we expect substantial speedups over spin-loop variants. %We analyze these effects in Section~\S\ref{sec:eval}.

% This work examines the impact of edge-centric and vertex-centric DAG traversal on both performance and convergence. Due to the non-deterministic order of atomic updates in edge-centric kernels, the solver with edge-centric kernels may converge in a number of iterations that is not same as the iteration count for vertex-centric kernels.
% %This means the iteration count may vary between successive runs, particularly when the convergence tolerance is high. 
% In contrast, vertex-centric kernels compute unknowns for each row serially, ensuring a deterministic order of computation.
% DAG-based solves on sparse matrices can achieve significantly higher bandwidth utilization than synchronization-free approaches~\cite{helal}, even for matrices without non-overlapping subdomains. By performing DAG traversal within subdomains and leveraging shared memory for vector data access, we expect substantial performance improvements over spin-loop-based variants. We evaluate the impact of these DAG-based optimizations in Section~\S\ref{sec:eval}.

\subsection{ILDU0 decomposition}
\label{ssec:ildu0}

In both synchronization-free and DAG-based lower solves (Algorithms~\ref{alg:lower_triangular_solve} and~\ref{alg:parallel_triangular_solve}), the updated value of the unknown for each row is divided by the diagonal element of that row. 
When a wavefront is assigned a row, only a representative thread from the wavefront performs the diagonal scaling step. 
Consequently, the scaling operation is carried out within a branch executed exclusively by the representative thread.

Branch divergence, introduced by this scaling process, can negatively impact kernel performance by causing other threads in the warp to stall.
To address this issue, we modify the ILU0 calculation by isolating diagonal entries from the lower and upper triangular matrices, as illustrated in Figure~\ref{fig:ildu} below. 
\begin{figure}[h]
\vspace{-6pt}
    \centering
    \includegraphics[width=0.45\textwidth]{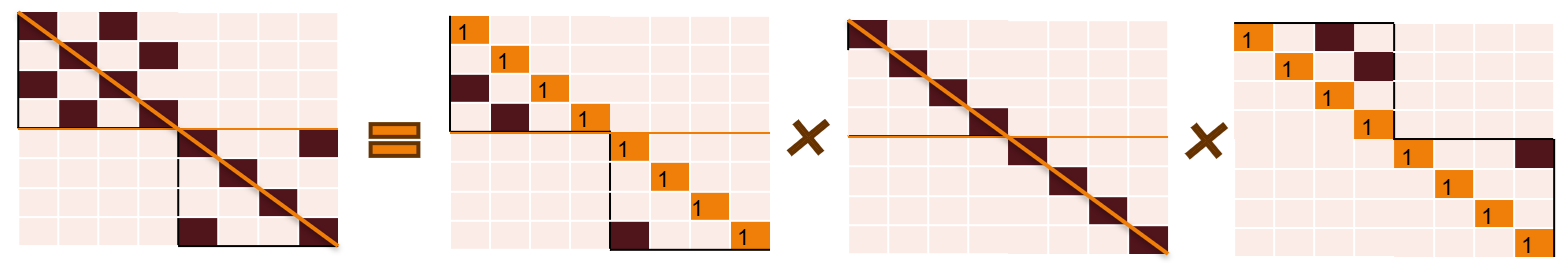}
    \caption{ILDU0 decomposition.}
    \label{fig:ildu}
\vspace{-6pt}
\end{figure}
Each row of the triangular matrix with non-unit diagonals and the input vector is divided by the corresponding diagonal value, ensuring the diagonal elements of the updated triangular matrix are all equal to one. This modification eliminates the need for diagonal scaling within a divergent branch.

Algorithm~\ref{alg:ILU0_ILDU0} shows the steps for computing the ILU0 and ILDU0 factorizations.
In our implementation, we use the \texttt{rocSPARSE} library~\cite{rocsparse} to generate the ILU0 decomposition of the reordered matrix with non-overlapping subdomains.
\texttt{rocSPARSE} generates ILU0 decomposition with one of the two triangular matrices having unit diagonals.
We then compute the ILDU0 decomposition by dividing the elements of the matrix with non-unit diagonals by its diagonal values, and explicitly store the inverse of the diagonal entries in a separate array.
Once the lower triangular solve is complete, the scaling operation for vector rows is performed. 
At this stage, threads can independently scale their respective vector elements, avoiding branch divergence.% and improving kernel performance.
% \begin{wrapfigure}{r}{0.55\columnwidth}
% \vspace{-10pt}
% \begin{minipage}{0.55\columnwidth}

\begin{algorithm}[t]
{\small
\caption{ILU0 and ILDU0 Factorizations~\cite{saad}}
\label{alg:ILU0_ILDU0}
\KwIn{Matrix $A$ of size $n \times n$}
\KwOut{For ILU0: $L$, $U$; \quad For ILDU0: $L$, $U$, inverse diagonal array \texttt{inv\_D}}

\vspace{1ex}
\textbf{ILU0:}\\
Initialize $L \gets I$, $U \gets A$\;

\For{$i \gets 1$ \KwTo $n$}{
    \For{$j \gets 1$ \KwTo $i-1$ \textbf{where} $A[i, j] \neq 0$}{
        $L[i, j] \gets U[i, j] / U[j, j]$\;
        \For{$k \gets j$ \KwTo $n$ \textbf{where} $A[i, k] \neq 0$}{
            $U[i, k] \gets U[i, k] - L[i, j] \cdot U[j, k]$\;
        }
    }
}

\vspace{1ex}
\textbf{ILDU0 (postprocessing):}\\
Initialize array \texttt{inv\_D} of length $n$\;

\For{$i \gets 1$ \KwTo $n$}{
    \texttt{inv\_D[$i$]} $\gets 1 / U[i, i]$\;
    \For{$j \gets 1$ \KwTo $i-1$ \textbf{where} $U[i, j] \neq 0$}{
        $U[i, j] \gets U[i, j] \cdot \texttt{inv\_D[$i$]}$\;
    }
}
}
\end{algorithm}

% \end{minipage}
% \vspace{-5pt}
% \end{wrapfigure}
\subsection{Kernel fusion}

Figure~\ref{fig:ildu} illustrates the application of the ILDU0 preconditioner on a matrix with two independent subdomains.
Since the subdomains do not overlap, all dependencies within a subdomain are confined to that subdomain.
%, with no external dependencies.
When thread blocks are assigned to subdomains, this structure provides a unique opportunity to fuse the lower triangular solve, diagonal scaling, and upper triangular solve into a single kernel. 
Thread block-wide synchronization is invoked after completing the lower triangular solve and after scaling to ensure the required results are available before proceeding to the next step.

The advantages of kernel fusion extend beyond reducing kernel launch overhead. 
Without fusion, each of the three steps (lower triangular solve, diagonal scaling, and upper triangular solve) would need to load vector data from global memory separately.
With fusion, the results of the lower triangular solve are directly updated in shared memory and are immediately available for the scaling step. 
Similarly, results of the scaling step are directly accessible for the upper triangular solve. This approach not only reduces the number of kernel invocations but also reduces the number of expensive global memory loads.

\subsection{Multi-GPU realization of ILDU0 application}
% As  dependencies within each subdomain are fully contained, our ILDU0 application strategy extends to multi-GPU systems. By evenly distributing subdomains across available GPUs—provided each subdomain is large enough to fully utilize a compute unit and the total number of subdomains exceeds the aggregate number of CUs—we can expect near-linear scaling.
As row dependencies for triangular solve kernels are confined within subdomains, our ILDU0 strategy extends naturally to multi-GPU systems. We expect near-linear scaling, provided subdomains are sufficiently large to utilize a compute unit and their total count exceeds the number of available CUs.

In this work, we evaluate the performance of a multi-GPU ILDU0 application, comprising the lower triangular solve, diagonal scaling, and upper triangular solve. Although the entire BiCGSTAB solver can, in principle, be parallelized across GPUs, steps such as sparse matrix-vector multiplication (SpMV) and norm evaluations would introduce inter-GPU communication. To isolate the performance characteristics of the ILDU0 component without communication-induced overheads, we restrict the scope of our multi-GPU evaluation to ILDU0. Full solver parallelization is left for future exploration.

In \S\ref{sec:eval}, we analyze the performance of synchronization-free and DAG-based triangular solves and evaluate the impact of domain decomposition on the overall performance of the BiCGSTAB solver.
\section{Performance evaluation}
\label{sec:eval}
%\atharva{TODO: Add the evaluation of latest cugraph jaccard}
%\atharva{1. VC legacy 2. Latest cuGraph 3. EC 4. EC + thread-dyn 5. EC + warp-dyn 6. EC + workgroup-dyn }

We conduct our experiments on a system with an AMD EPYC\textsuperscript{TM} 7763 64-core CPU and eight AMD Instinct\textsuperscript{TM} MI210 GPUs, using the \texttt{hipcc} compiler along with \texttt{rocSPARSE} and \texttt{rocBLAS} from ROCm 6.1.2. Table~\ref{tab:inputs} summarizes the sparse datasets used. \textbf{Laplacian1} and \textbf{Laplacian2} are $3 \times 3$ block sparse matrices (BSR) derived from 3D Laplacian discretizations. All matrices other than the variants of Laplacian discretizations are in CSR format. Next, \textbf{parabolic\_fem}, taken from the SuiteSparse Matrix Collection~\cite{suitesparse,suitesparseweb}, is generated from diffusion-convection-reaction equations, while \textbf{spe10} is based on permeability data from an oil reservoir simulation~\cite{spe10,spe10paper}.
Lastly, \textbf{rhd}, \textbf{rhd-3T}, and \textbf{oil} are provided by Zong et al.~\cite{zong,zenodo}.

\begin{figure}[h]
    \vspace{-4pt}
    \centering
    \begin{subfigure}[b]{0.22\textwidth}
        \centering
        \includegraphics[width=\textwidth]{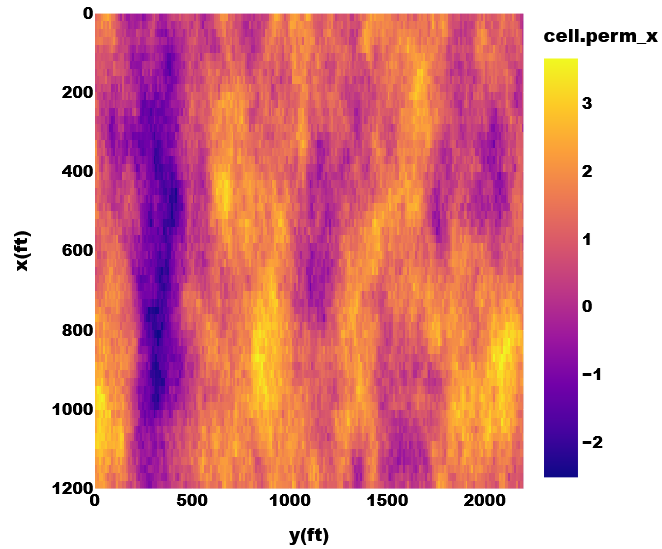}
        \caption{Layer 0}
        \label{fig:layer0}
    \end{subfigure}
   \begin{subfigure}[b]{0.22\textwidth}
        \centering
        \includegraphics[width=\textwidth]{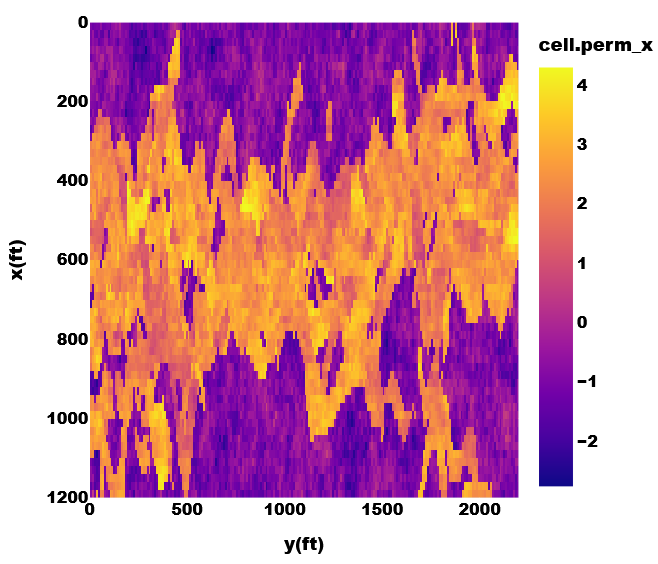}
        \caption{Layer 80}
        \label{fig:layer80}
    \end{subfigure}
    % \begin{subfigure}[b]{0.24\textwidth}
    %     \centering
    %     \includegraphics[width=\textwidth]{figures/perm40.png}
    %     \caption{Layer 40}
    %     \label{fig:layer40}
    % \end{subfigure}
    % \begin{subfigure}[b]{0.24\textwidth}
    %     \centering
    %     \includegraphics[width=\textwidth]{figures/perm80.png}
    %     \caption{Layer 80 }
    %     \label{fig:layer80}
    % \end{subfigure}
    \caption{Variation in permeability of spe10~\cite{spe10,spe10paper} layers.}
 \label{fig:spe10perm}
\end{figure}

\textbf{spe10} is an important stress test for our approach, as it originates from a porous medium with significant permeability variation (Figure~\ref{fig:spe10perm} above), both within (Figure~\ref{fig:layer0}) and across layers (Figures~\ref{fig:layer0}--\ref{fig:layer80}). These variations lead to a poorly conditioned system that requires more solver iterations. 
We use a convergence tolerance of \textbf{10\textsuperscript{-8}} for all matrices in Table~\ref{tab:inputs}, except \textbf{spe10}, where we relax it to \textbf{10\textsuperscript{-6}}.

%poorly conditioned sytems: Ratio bewteen smallest and largest eigenvalue are likely closer to zero.
%perhaps worth mentioning what a poorly conditioned matrix is and explain why poorly conditioned matrices are not ideal

%\atharva{TODO: Add EVAL}
We categorize the performance evaluation into four parts: (1) performance of \texttt{rocSPARSE} and our triangular solve kernels, (2) performance of the preconditioner application, including the lower triangular solve with diagonal scaling and the upper triangular solve,  (3) performance evaluation of the iterative solver with and without domain decomposition, and (4) multi-GPU implementation of ILDU0 preconditioner application on subdomains.
\begin{table}[b]
\caption{Matrices used in the evaluation.}
\label{tab:inputs}
\vspace*{-6pt}
\centering
%\huge
%\scalebox{0.5}
\resizebox{\columnwidth}{!}
{
\begin{tabular}{|l|l|l|l|l|l|}
\hline
Matrix         & \begin{tabular}[c]{@{}l@{}}Data \\ structure\end{tabular} & Description                                                                                                             & \begin{tabular}[c]{@{}l@{}}\#Rows\\ (M=millions)\end{tabular} & \begin{tabular}[c]{@{}l@{}}\#Nonzeros\\ (M=millions)\end{tabular} & \begin{tabular}[c]{@{}l@{}}Real \\ world?\end{tabular} \\ \hline
Laplacian1     & BSR  & \begin{tabular}[c]{@{}l@{}}Discretization of\\ Laplacian PDE\end{tabular}                                               & 6.3M                                                          & 131.2M                                                             & No                                                     \\ \hline
Laplacian2     & BSR  & \begin{tabular}[c]{@{}l@{}}Discretization of\\ Laplacian PDE\end{tabular}                                               & 12.6M                                                         & 262.8M                                                            & No                                                     \\ \hline
parabolic\_fem & CSR  & \begin{tabular}[c]{@{}l@{}}Diffusion-\\ convection reaction\end{tabular}                                                & 0.5M                                                          & 3.6M                                                              & Yes                                                    \\ \hline
spe10          & CSR  & \begin{tabular}[c]{@{}l@{}}Reservoir \\ simulation on\\ geocellular models\end{tabular}                                 & 1.1M                                                          & 7.7M                                                              & Yes                                                    \\ \hline
rhd            & CSR  & \begin{tabular}[c]{@{}l@{}}Radiation\\ hydrodynamics\end{tabular}                                                       & 2.1M                                                          & 14.5M                                                             & Yes                                                    \\ \hline
rhd-3T         & CSR  & \begin{tabular}[c]{@{}l@{}}Radiation \\ hydrodynamics\\ 3T - 3 components,\\ (radiation, electron, \\ ion)\end{tabular} & 6.3M                                                          & 52.1M                                                             & Yes                                                    \\ \hline
oil            & CSR  & \begin{tabular}[c]{@{}l@{}}Petroleum reservoir \\ simulation\end{tabular}                                               & 31.5M                                                         & 219.0M                                                            & Yes                                                    \\ \hline
\end{tabular}
}
\end{table} 

%\atharva{TODO:Add tolerances in input  data table}

\subsection{Evaluation of lower triangular solves}
We evaluate the performance of the following implementations of triangular solves on matrices with non-overlapping subdomains:
\begin{itemize}[leftmargin=*]

  \item \textbf{\texttt{rocSPARSE} lower triangular solve}: Baseline kernel using synchronization free solves from the \texttt{rocSPARSE} library. 
\end{itemize}
  \textbf{The following implementations were developed in this work:}
  \begin{itemize}[leftmargin=*]
  \item \textbf{\texttt{spin\_loop\_lds}}: This is our extension to \texttt{rocSPARSE} triangular solves. We assign a subdomain to each thread block, and store partial sums per row in shared memory. The solve proceeds in a synchronization-free manner.

  \item \textbf{\texttt{dag\_ec\_no\_lds}}: We construct a DAG for each subdomain and assign it to a thread block. The kernel performs edge-centric traversal using global memory for vector access to isolate DAG-related benefits.

  \item \textbf{\texttt{dag\_ec\_lds}}: We extend \texttt{dag\_ec\_no\_lds} by loading vector data into shared memory before performing the DAG-based triangular solve.

  \item \textbf{\texttt{dag\_ec\_lds\_ud}}: ILDU0 variant of \texttt{dag\_ec\_lds}. We assume a unit diagonal (UD) and defer diagonal scaling until after the triangular solve, unlike ILU0 where scaling follows each row's off-diagonal updates.
\end{itemize}

\begin{table}[]

\caption{Runtime of single invocation of lower triangular solve kernels (milliseconds).}
\label{tab:ltseval}
\vspace*{-6pt}
\centering
\huge
%\scalebox{0.45}
\resizebox{\columnwidth}{!}
{
\begin{tabular}{l|lllll|}
\cline{2-6}
                                     & \multicolumn{5}{c|}{Implementation}                                                                                                                                              \\ \hline
\multicolumn{1}{|l|}{Input}          & \multicolumn{1}{l|}{rocsparse\_lower} & \multicolumn{1}{l|}{spin\_loop\_lds} & \multicolumn{1}{l|}{dag\_ec\_no\_lds} & \multicolumn{1}{l|}{dag\_ec\_lds} & dag\_ec\_lds\_ud \\ \hline
\multicolumn{1}{|l|}{laplacian1}     & \multicolumn{1}{l|}{7.221}                 & \multicolumn{1}{l|}{5.912}           & \multicolumn{1}{l|}{2.363}            & \multicolumn{1}{l|}{0.907}        & 0.793            \\ \hline
\multicolumn{1}{|l|}{laplacian2}     & \multicolumn{1}{l|}{15.109}                & \multicolumn{1}{l|}{12.487}           & \multicolumn{1}{l|}{4.576}            & \multicolumn{1}{l|}{1.834}        & 1.579            \\ \hline
\multicolumn{1}{|l|}{parabolic\_fem} & \multicolumn{1}{l|}{0.556}                 & \multicolumn{1}{l|}{0.858}           & \multicolumn{1}{l|}{0.182}            & \multicolumn{1}{l|}{0.060}        & 0.055            \\ \hline
\multicolumn{1}{|l|}{spe10}          & \multicolumn{1}{l|}{1.612}                 & \multicolumn{1}{l|}{2.244}           & \multicolumn{1}{l|}{0.537}            & \multicolumn{1}{l|}{0.185}        & 0.178            \\ \hline
\multicolumn{1}{|l|}{rhd}            & \multicolumn{1}{l|}{3.192}                 & \multicolumn{1}{l|}{3.304}           & \multicolumn{1}{l|}{1.387}            & \multicolumn{1}{l|}{0.285}        & 0.268            \\ \hline
\multicolumn{1}{|l|}{rhd-3T}         & \multicolumn{1}{l|}{8.044}                 & \multicolumn{1}{l|}{8.298}           & \multicolumn{1}{l|}{2.488}            & \multicolumn{1}{l|}{0.778}        & 0.737            \\ \hline
\multicolumn{1}{|l|}{oil}            & \multicolumn{1}{l|}{48.781}                & \multicolumn{1}{l|}{40.939}          & \multicolumn{1}{l|}{16.623}           & \multicolumn{1}{l|}{4.393}        & 4.180            \\ \hline
\end{tabular}
}
\end{table}
%Table~\ref{tab:ltseval} shows..

\begin{figure}[b] 
%\vspace{-10}
\centering
\includegraphics[width=\columnwidth]%{figures/lts_speedups.png}
{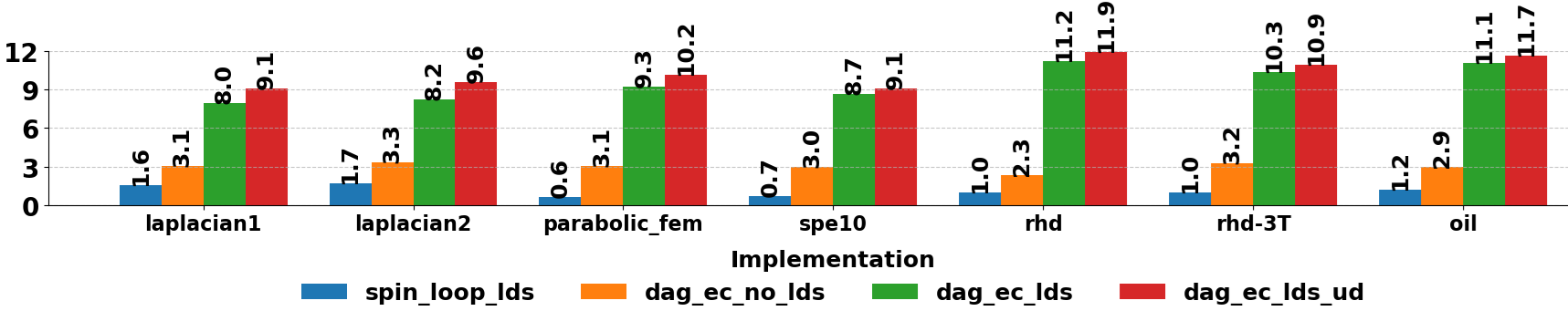}
\vspace*{-10pt}
\caption{Speedup over \texttt{rocSPARSE}~\cite{rocsparse} lower triangular solves.}
\label{fig:ltsspeedup}
%\vspace{-10}
\end{figure}

\noindent Table~\ref{tab:ltseval} presents the runtimes of the lower triangular solve kernels. The spin-loop-based kernel with shared memory usage outperforms the baseline \texttt{rocSPARSE} kernels for larger graphs (e.g., \textit{laplacian1}, \textit{laplacian2}, \textit{oil}) by a factor of up to $1.7 \times$.
Overall, DAG-based approaches demonstrate significant performance improvements over synchronization-free variants. 
Notably, the \texttt{dag\_ec\_lds\_ud} kernel outperforms the \texttt{rocSPARSE} kernel by a factor of up to $11.9 \times$.
Figure~\ref{fig:ltsspeedup} below shows the speedup of our implementations over the \texttt{rocSPARSE} triangular solve kernel, highlighting the performance gains achieved with our approach.
Next, we evaluate the impact of DAG-based
solves on the performance of preconditioner application kernels and explain the observed trends using memory bandwidth and VALU utilization metrics.

% Table~\ref{tab:ltseval} presents the runtimes of the lower triangular solve kernels. The spin-loop-based kernel with shared memory usage outperforms the baseline \texttt{rocSPARSE} kernels for larger graphs (e.g., \textit{laplacian1}, \textit{laplacian2}, \textit{oil}) by a factor of up to $1.2 \times$. DAG-based approaches demonstrate significant performance improvements over synchronization-free variants. 
% Notably, the \texttt{dag\_ec\_lds\_ud} kernel outperforms the \texttt{rocSPARSE} kernel by a factor of up to $11 \times$. 

% Figure~\ref{fig:ltsspeedup} shows the speedup of our implementations over the \texttt{rocSPARSE} triangular solve kernel, highlighting the performance gains achieved with our approach.

% Next, we explore the performance of the application of the preconditioner generated using both the ILU0 and ILDU0 approaches.

\subsection{Evaluation of the application of preconditioner}
We evaluate four implementations of the preconditioner application using lower and upper triangular solves:
\begin{itemize}[leftmargin=*]
  \item \textbf{\texttt{rocSPARSE\_ILU0}}: Uses two \texttt{rocSPARSE} kernels~\cite{rocsparse} for the lower and upper triangular solves.
\end{itemize}
\textbf{The following implementations were developed in this work:}
\begin{itemize}[leftmargin=*]
  \item \textbf{\texttt{dag\_ec\_ILD\_U0}}: Edge-centric DAG-based ILU0 with separate kernels for lower (LD) and upper solves. Both load vector data into shared memory.

  \item \textbf{\texttt{dag\_vc\_ILDU0\_fused}}: Vertex-centric DAG traversal for ILDU0 preconditioner, implemented as a single kernel for lower solver, diagonal scaling, and upper triangular solve. Vector data is loaded into shared memory once, with subsequent steps updating it in-place.

  \item \textbf{\texttt{dag\_ec\_ILDU0\_fused}}: Uses edge-centric DAG traversal.
\end{itemize}

% 1. \texttt{rocSPARSE\_ILU0}: This implementation consists of two kernels from \texttt{rocSPARSE} library~\cite{rocsparse}: one for the lower triangular solve and another for the upper triangular solve.

% 2. \texttt{dag\_ec\_ILD\_U0}: This kernel employs a DAG-based approach for matrix ILU0 decomposition. Two kernels are designed: one for computing the lower triangular solve while scaling the diagonal (LD) and another for performing the upper triangular solve. Vector data is loaded into shared memory in both kernels.

% 3. \texttt{dag\_vc\_ILDU0\_fused}: This is the vertex-centric kernel for DAG traversal on the ILDU0 preconditioner. It is implemented as a single kernel, with vector data loaded into shared memory only once, before the start of the lower triangular solve. Inputs for subsequent steps (diagonal scaling and upper triangular solve) are updated directly in shared memory at the end of the previous step.

% 4. \texttt{dag\_ec\_ILDU0\_fused}: Similar in concept to 
% \texttt{dag\_vc\_ILDU0\_fused} in terms of shared memory usage and single kernel invocation. The primary difference is that DAG traversal is performed in an edge-centric manner.

\noindent Table~\ref{tab:ldueval} below presents the runtimes of the preconditioner application kernels on the AMD Instinct MI210 GPU. Figure~\ref{fig:overall} below shows the speedup of our implementations over the baseline \texttt{rocSPARSE\_ILU0} implementation. The edge-centric kernel with fused ILDU0 (\texttt{dag\_ec\_ILDU0\_fused}) outperforms the \texttt{rocSPARSE\_ILU0} implementation by 10.7$\times$.
%a factor of 10.7 (geometric mean). 
\begin{table}[h]
\caption{Runtime (milliseconds) of the single invocation of preconditioner using ILU0/ILDU0.}
\label{tab:ldueval}
\vspace*{-6pt}
\centering
\huge
%\scalebox{0.45}
\resizebox{\columnwidth}{!}
{
\begin{tabular}{l|llll|}
\cline{2-5}
                                     & \multicolumn{4}{c|}{Implementation}                                                                                                              \\ \cline{2-5} 
                                     & \multicolumn{1}{l|}{rocsparse\_ILU0} & \multicolumn{1}{l|}{dag\_ec\_ILD\_U0} & \multicolumn{1}{l|}{dag\_vc\_ILDU0\_fused} & dag\_ec\_ILDU0\_fused \\ \hline
\multicolumn{1}{|l|}{laplacian1}     & \multicolumn{1}{l|}{14.33}           & \multicolumn{1}{l|}{1.68}             & \multicolumn{1}{l|}{2.21}                  & 1.53                 \\ \hline
\multicolumn{1}{|l|}{laplacian2}     & \multicolumn{1}{l|}{29.12}           & \multicolumn{1}{l|}{3.40}             & \multicolumn{1}{l|}{4.51}                  & 3.16                 \\ \hline
\multicolumn{1}{|l|}{parabolic\_fem} & \multicolumn{1}{l|}{1.11}            & \multicolumn{1}{l|}{0.11}             & \multicolumn{1}{l|}{0.18}                  & 0.10                 \\ \hline
\multicolumn{1}{|l|}{spe10}          & \multicolumn{1}{l|}{3.22}            & \multicolumn{1}{l|}{0.36}             & \multicolumn{1}{l|}{0.68}                  & 0.34                 \\ \hline
\multicolumn{1}{|l|}{rhd}            & \multicolumn{1}{l|}{6.38}            & \multicolumn{1}{l|}{0.55}             & \multicolumn{1}{l|}{1.03}                  & 0.51                 \\ \hline
\multicolumn{1}{|l|}{rhd-3T}         & \multicolumn{1}{l|}{16.09}           & \multicolumn{1}{l|}{1.52}             & \multicolumn{1}{l|}{2.34}                  & 1.44                 \\ \hline
\multicolumn{1}{|l|}{oil}            & \multicolumn{1}{l|}{97.56}           & \multicolumn{1}{l|}{8.57}             & \multicolumn{1}{l|}{12.68}                 & 7.60                 \\ \hline
\end{tabular}
}
\end{table} 
%\atharva{explain why ? nnzs>#rows more work to be doen in parallel}
%  We present the average read bandwidth of the preconditioner application kernels in Figure~\ref{fig:readbw}. Edge-centric variants consistently achieve higher read bandwidth compared to \texttt{rocSPARSE} and vertex-centric kernels. This is expected, as the number of nonzeros in a graph typically exceeds the number of rows. When parallelism is extracted from nonzeros, threads can issue read requests for the same row in parallel, in contrast to the serialized reads per row in the vertex-centric approach.

\begin{figure}[h] 
\centering
\includegraphics[width=\columnwidth]%{figures/read_bandwidth.png}
{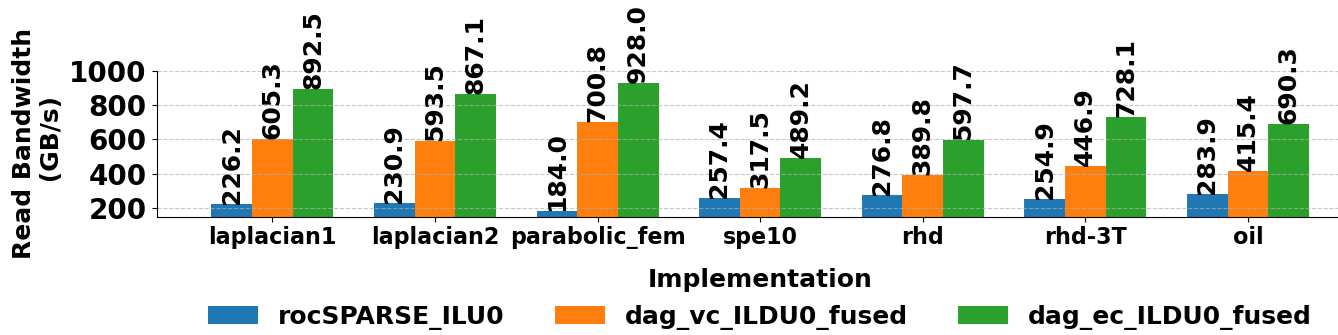}
\vspace{-12pt}
\caption{Read bandwidth (GBPS): ILU0 application kernels.}
\label{fig:readbw}
\vspace{-12pt}
\end{figure}

\begin{figure}[htb] 
\centering
\includegraphics[width=\columnwidth]{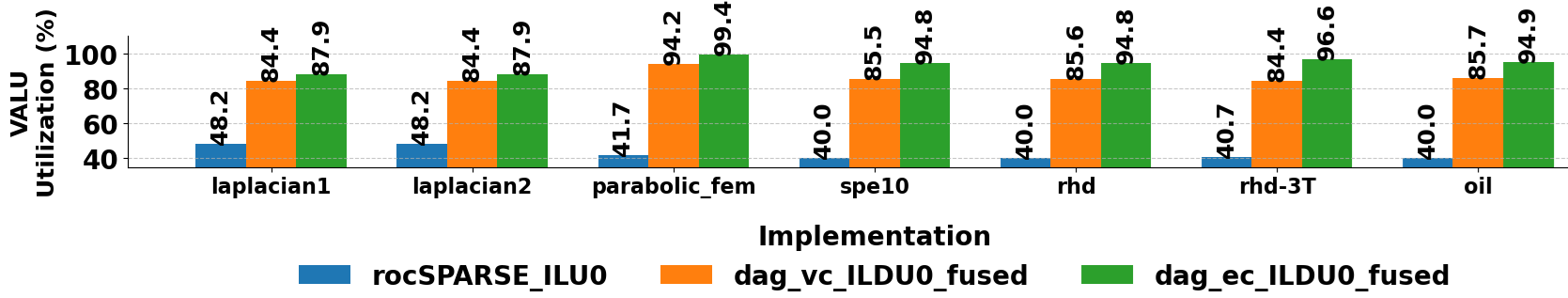}
\vspace{-12pt}
\caption{Vector ALU utilization for ILU0 application kernels.}
\label{fig:valu}
\vspace{-10pt}
\end{figure}

% In Figure~\ref{fig:valu}, we analyze thread activity using the vector ALU utilization (\texttt{VALUUtilization}) metric, collected with the \texttt{rocProf} profiler~\cite{rocprof}. A \texttt{VALUUtilization} value of 100 indicates that all vector ALU threads of a wavefront remain active for the duration the wavefront executes the instructions. The edge-centric kernel demonstrates higher thread activity compared to \texttt{rocSPARSE\_ILU0}.
Figure~\ref{fig:readbw} shows the average read bandwidth of preconditioner application kernels. Edge-centric variants consistently outperform \texttt{rocSPARSE} and vertex-centric kernels, as parallelism over nonzeros allows concurrent row accesses, unlike the serialized reads in vertex-centric solves.
Figure~\ref{fig:valu} reports vector ALU utilization (\texttt{VALUUtilization}) collected via \texttt{rocProf}~\cite{rocprof}. A VALUUtilization value of 100 indicates full thread activity across a wavefront. Edge-centric kernels exhibit significantly higher vector ALU utilization than \texttt{rocSPARSE\_ILU0}.

Overall, the performance improvements of DAG-based solves over \texttt{rocSPARSE} kernels stem from three key factors: 
(1) the use of shared memory instead of global memory for irregular vector reads and writes, 
(2) higher bandwidth utilization enabled by explicit dependency tracking and elimination of inter-subdomain busy-waiting/synchronization, and
(3) the elimination of inter-subdomain dependencies, which enables kernel fusion for lower and upper triangular solves.

\begin{figure}[h] 
%\vspace{-10}
\centering
\includegraphics[width=\columnwidth]%{figures/ildu_new3.png}
{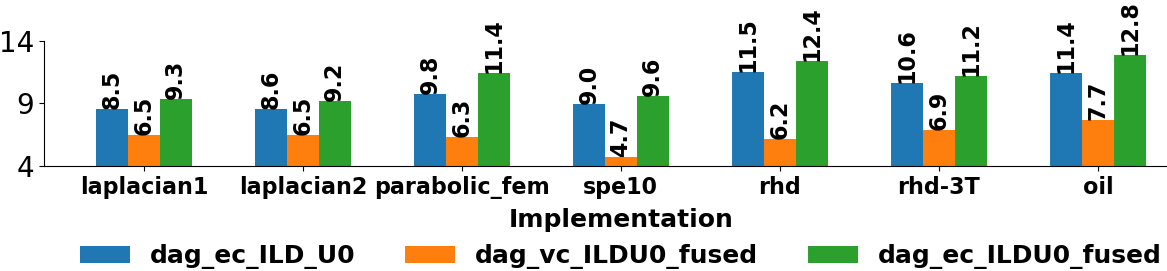}
\vspace{-12pt}
\caption{Speedup over ILU0 application with rocSPARSE~\cite{rocsparse} triangular solves.}
\label{fig:overall}
\end{figure}
\begin{figure}[h] 
\centering
\includegraphics[width=\columnwidth]%{figures/speedupoverall.png}
{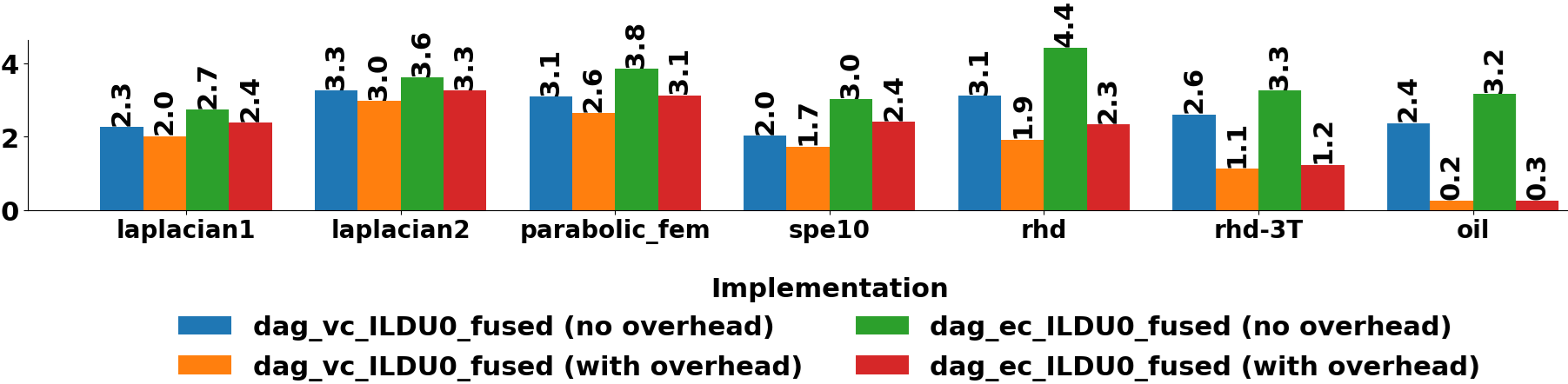}
\caption{Speedup over reference solver implementation~\cite{opm} with \texttt{rocSPARSE}~\cite{rocsparse} triangular solves.}
\vspace{-12pt}
\label{fig:overallsolver}
\end{figure}
\begin{table}[!htbp]
\vspace*{6pt}
\caption{Impact of partitioning.}
\label{tab:partitions}
\vspace*{-6pt}
\centering
\huge
%\scalebox{0.45}
\resizebox{\columnwidth}{!}
{
\begin{tabular}{|l|l|l|l|l|l|l|}
\hline
matrix                                                   & \begin{tabular}[c]{@{}l@{}}partitioning \\ algorithm\end{tabular} & \begin{tabular}[c]{@{}l@{}}rows per\\ subdomain\end{tabular} & \begin{tabular}[c]{@{}l@{}}number\\ of\\ subdomains\end{tabular} & \# nonzeros & \begin{tabular}[c]{@{}l@{}}\# nonzeros\\ post \\ decomposition\end{tabular} & \begin{tabular}[c]{@{}l@{}}nonzeros \\ dropped\\ (\%)\end{tabular} \\ \hline
laplacian1                                               & \begin{tabular}[c]{@{}l@{}}geometric \\ cuts\end{tabular}      & 2048                                                         & 1024                                                             & 131,235,840 & 122,683,392                                                                & 6.52                                                               \\ \hline
laplacian2                                               & \begin{tabular}[c]{@{}l@{}}geometric \\ cuts\end{tabular}      & 2048                                                         & 2048                                                             & 262,766,592 & 245,366,784                                                                & 6.62                                                               \\ \hline
\begin{tabular}[c]{@{}l@{}}parabolic\\ -fem\end{tabular} & metis                                                          & 8192                                                         & 61                                                               & 3,674,625   & 3,600,658                                                                  & 2.01                                                               \\ \hline
spe10                                                    & metis                                                          & 8192                                                         & 137                                                              & 7,780,000   & 7,403,366                                                                  & 4.84                                                               \\ \hline
rhd                                                      & metis                                                          & 8192                                                         & 256                                                              & 14,581,760  & 13,831,024                                                                 & 5.15                                                               \\ \hline
rhd-3T                                                   & metis                                                          & 8192                                                         & 768                                                              & 52,133,888  & 48,659,014                                                                 & 6.67                                                               \\ \hline
oil                                                      & metis                                                          & 8192                                                         & 3841                                                             & 219,046,528 & 207,390,544                                                                & 5.32                                                               \\ \hline
\end{tabular}
}
\end{table}
\vspace*{-6pt}

\begin{figure*}[h]
    \centering
    \begin{subfigure}[b]{0.24\textwidth}
        \centering
        \includegraphics[width=\textwidth]{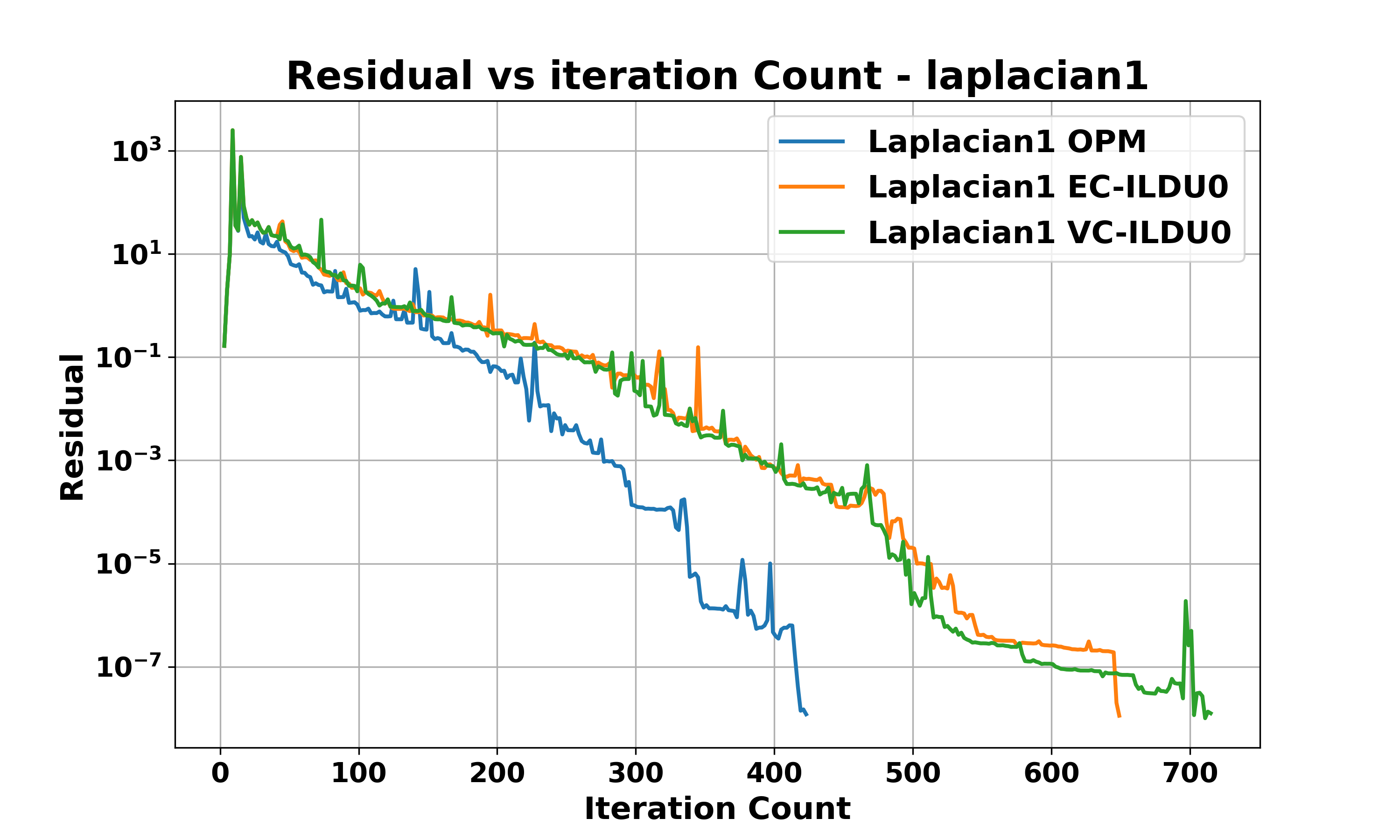}
        \caption{laplacian1}
        \label{fig:l2}
    \end{subfigure}
   \begin{subfigure}[b]{0.24\textwidth}
        \centering
        \includegraphics[width=\textwidth]{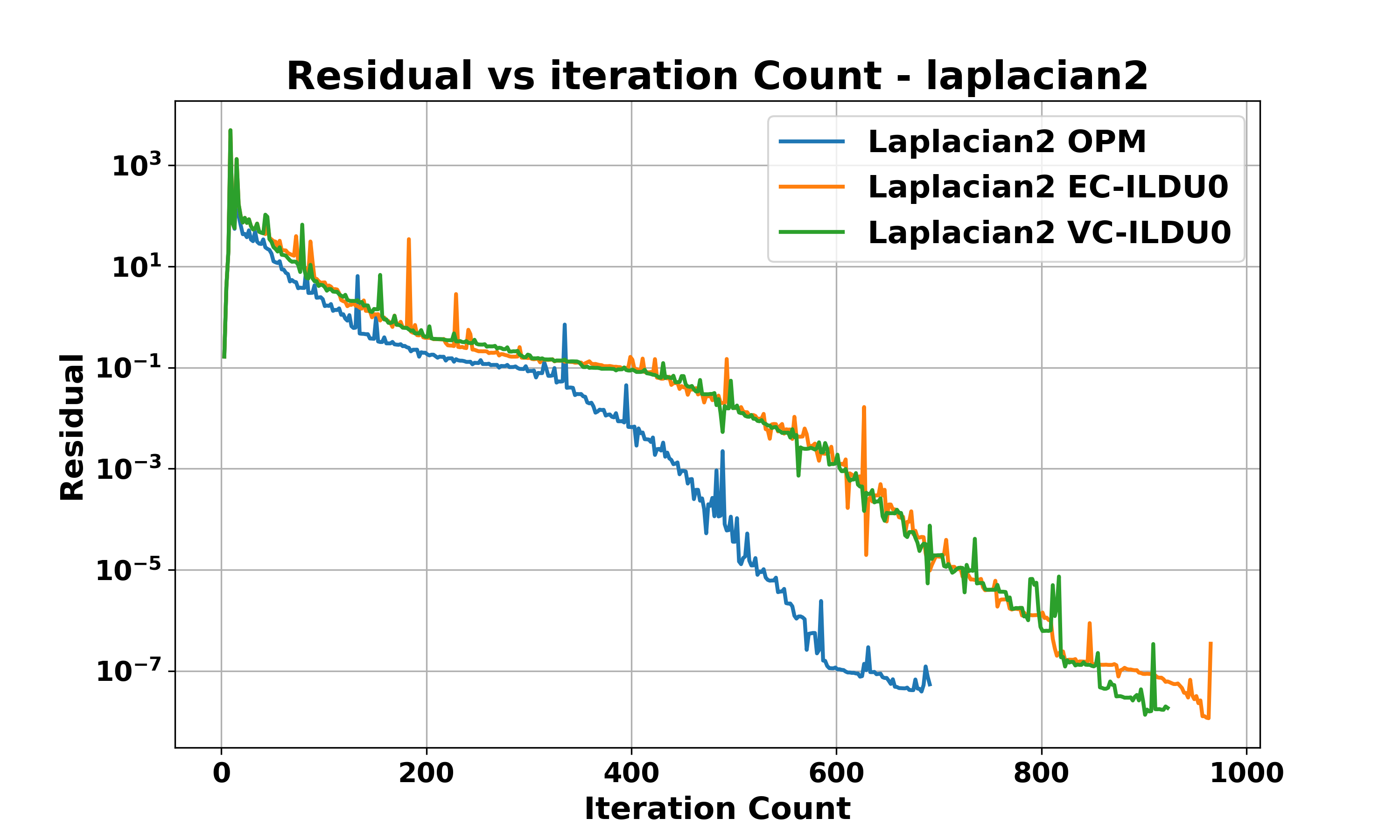}
        \caption{laplacian2}
        \label{fig:l2}
    \end{subfigure}
    \begin{subfigure}[b]{0.24\textwidth}
        \centering
        \includegraphics[width=\textwidth]{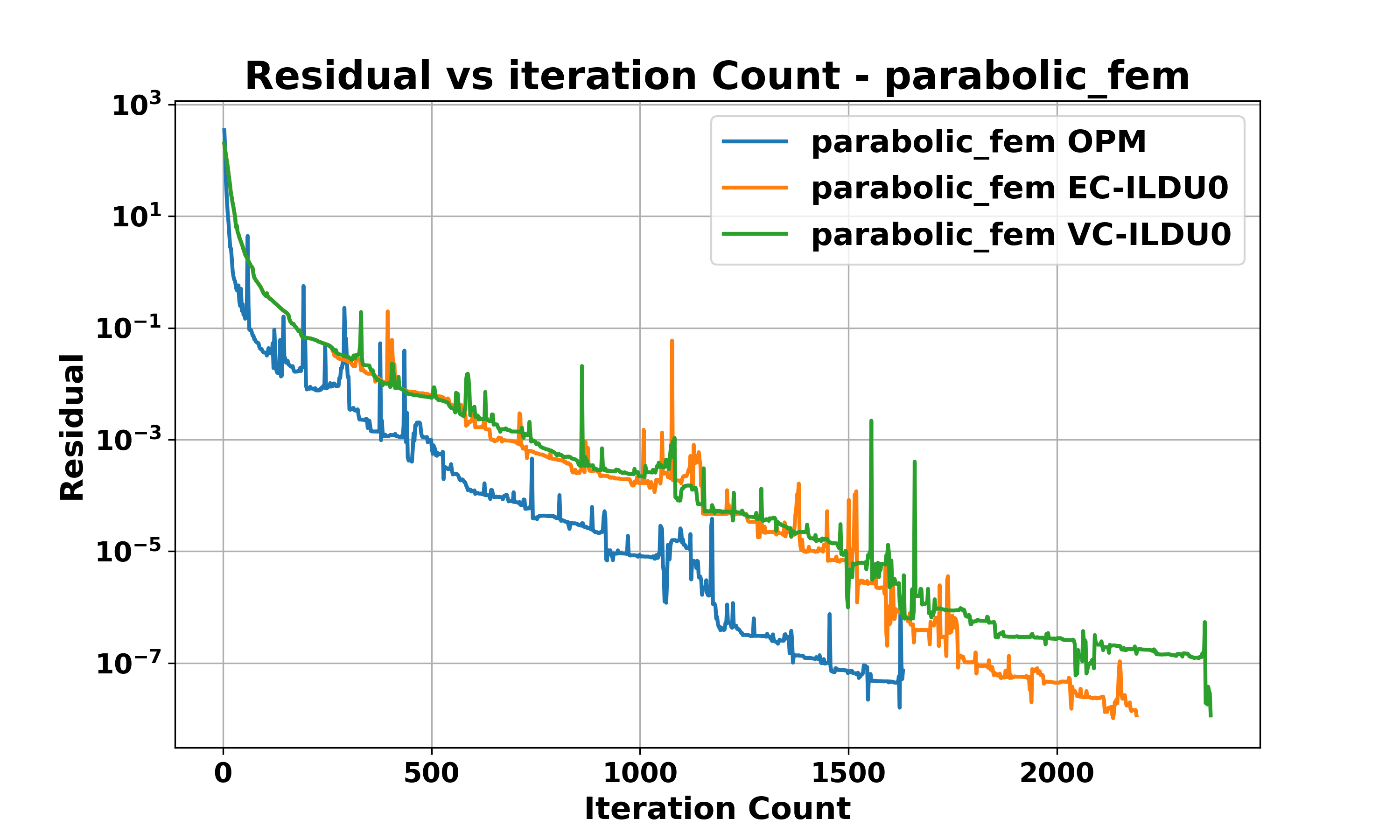}
        \caption{parabolic fem}
        \label{fig:pfem}
    \end{subfigure}
    \begin{subfigure}[b]{0.24\textwidth}
        \centering
        \includegraphics[width=\textwidth]{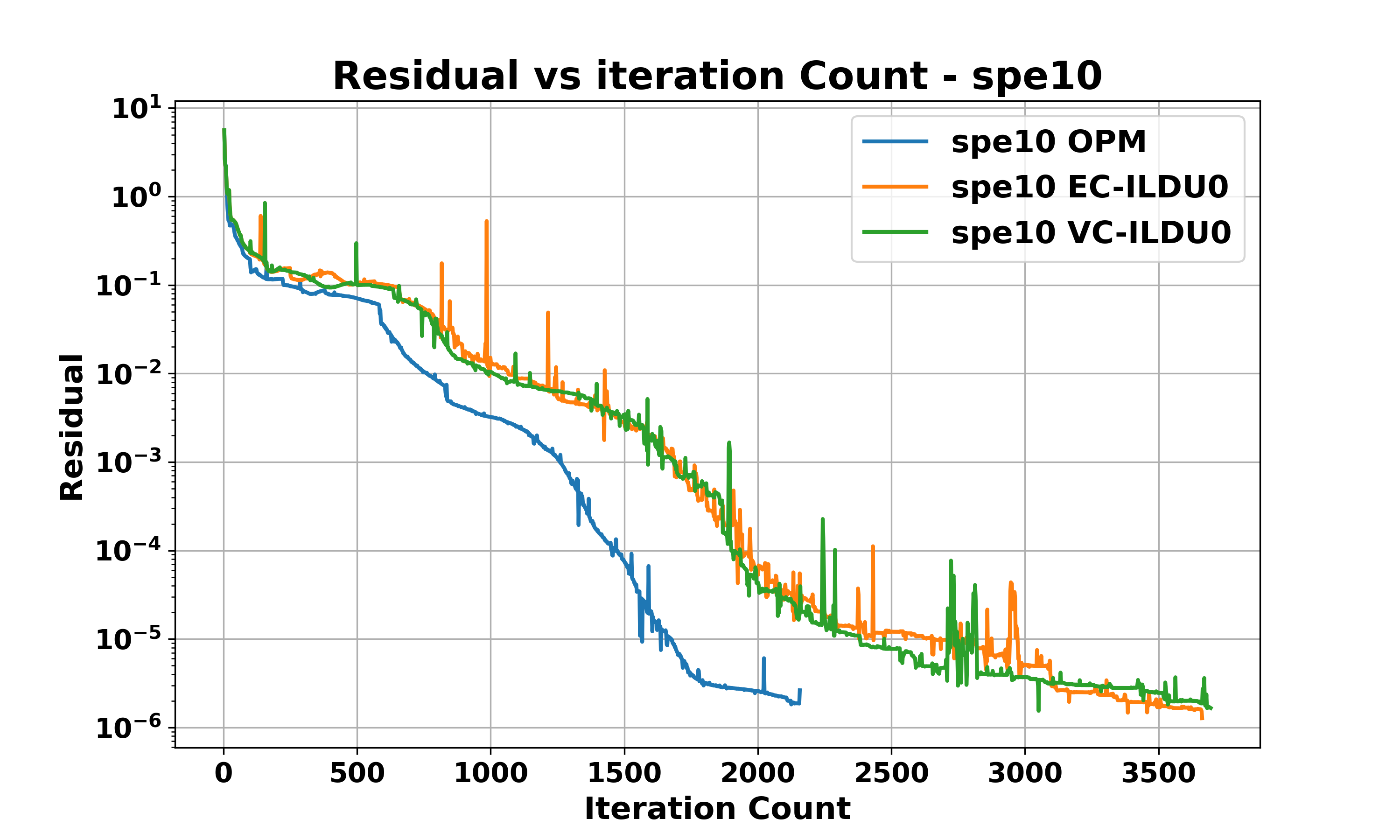}
        \caption{spe10}
        \label{fig:spe10}
    \end{subfigure}
    \begin{subfigure}[b]{0.24\textwidth}
        \centering
        \includegraphics[width=\textwidth]{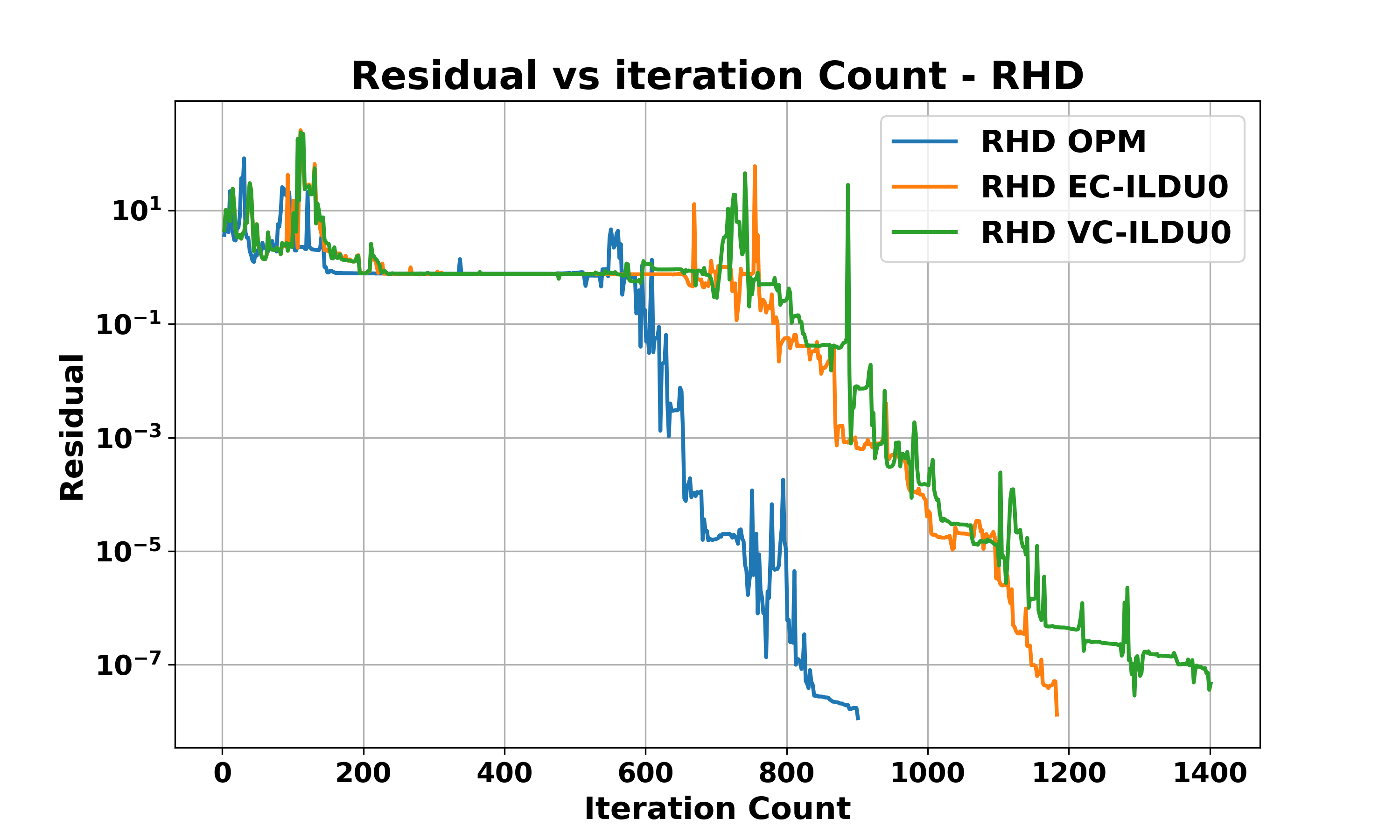}
        \caption{rhd}
        \label{fig:rhd}
    \end{subfigure}
    \begin{subfigure}[b]{0.24\textwidth}
        \centering
        \includegraphics[width=\textwidth]{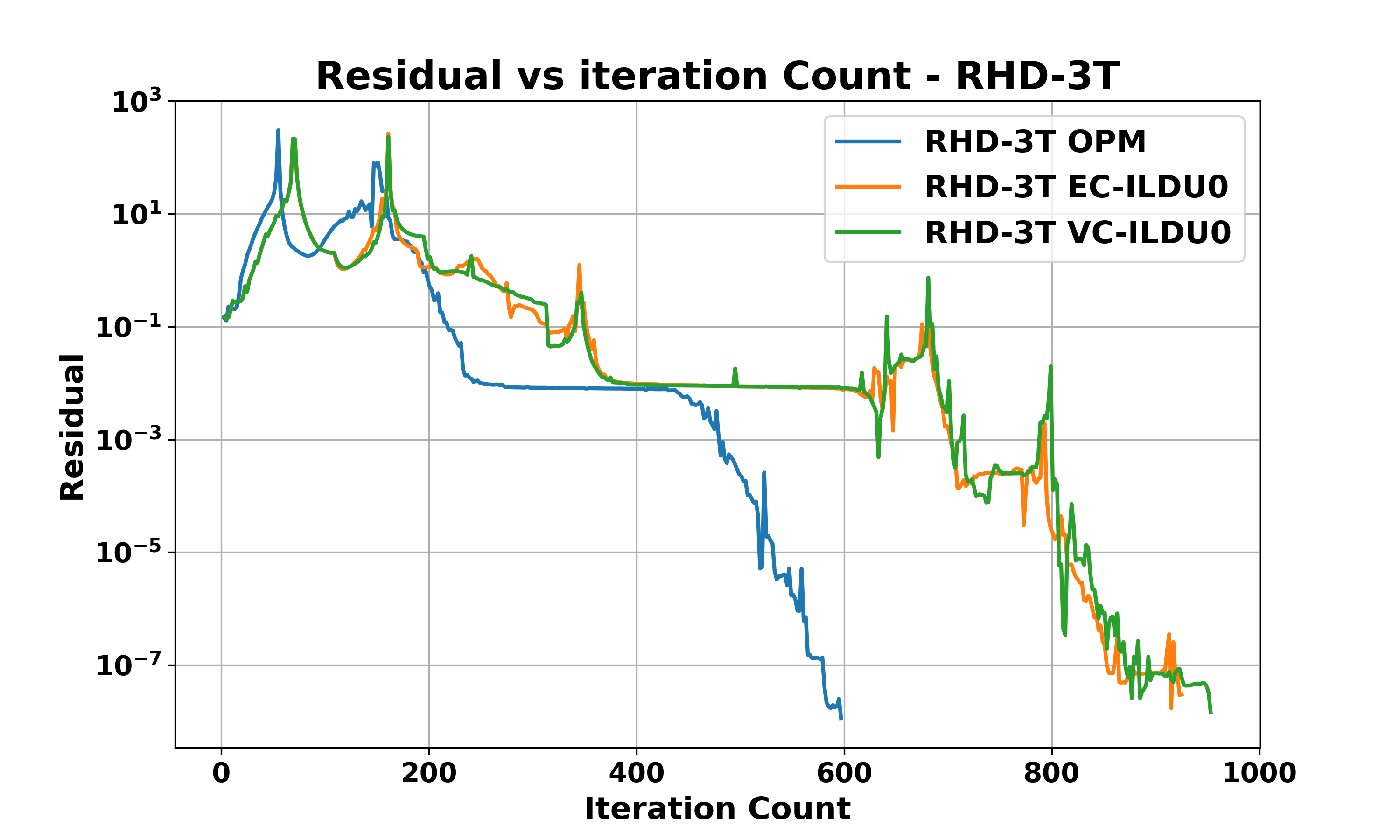}
        \caption{rhd-3T }
        \label{fig:rhd3t}
    \end{subfigure}
      \begin{subfigure}[b]{0.22\textwidth}
        \centering
        \includegraphics[width=\textwidth]{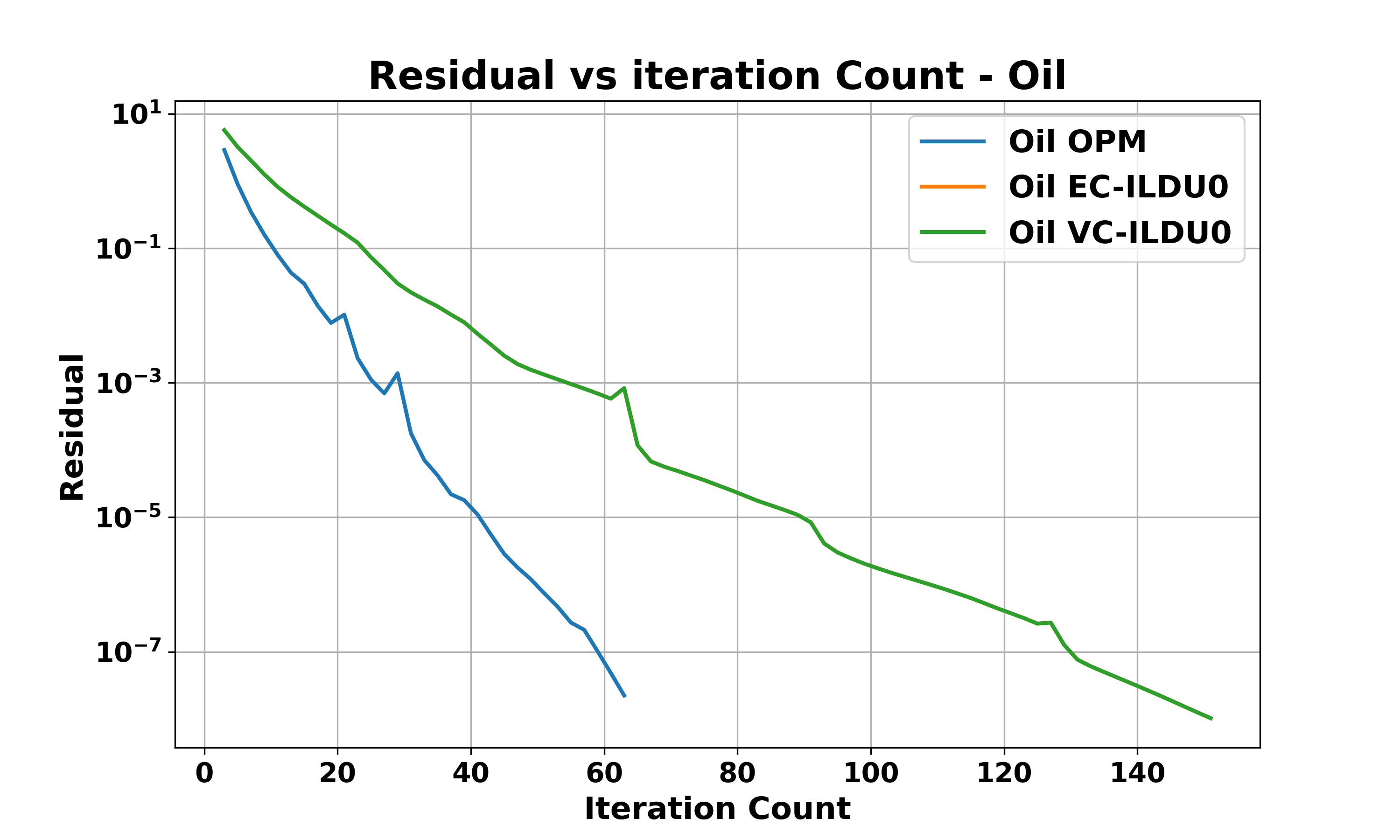}
        \caption{oil }
        \label{fig:oil}
    \end{subfigure}

    \caption{Convergence of solver with \texttt{rocSPARSE}, dag\_ec\_ILDU0\_fused, and dag\_vc\_ILDU0\_fused kernels for triangular solves.}
 \label{fig:resplots}
 \vspace{-10pt}
\end{figure*}

\subsection{Evaluation of solver runtime}

Table~\ref{tab:partitions} below shows the impact of domain decomposition on the number of nonzeros in the matrix. Domain decomposition reduces the number of nonzeros in the preconditioner matrix by 2-6.5\%.

For the \textit{laplacian1} and \textit{laplacian2} inputs, we use geometric cut partitioning to divide the matrix into uniform subdomains, each containing 2048 block rows (equivalent to 6144 CSR rows per subdomain). For real-world inputs, we use \texttt{METIS}~\cite{metis} to partition the matrix. We ensure that the partitions have a uniform size of 8192 rows by manually balancing the partition sizes if \texttt{METIS} does not produce uniform partitions.

Tables~\ref{tab:eceval} and~\ref{tab:vceval} present the runtimes of the BiCGTSAB solver with dag\_ec\_ILDU0\_fused and dag\_vc\_ILDU0\_fused, respectively, along with the iteration count and total overhead associated with domain decomposition. Solvers with domain decomposition require more iterations to converge but outperform solvers with \texttt{rocSPARSE} triangular solves.
%Table~\ref{tab:partitions}
%Table~\ref{tab:eceval}
%Table~\ref{tab:vceval}
%Our implementation achieves performance improvements over the baseline for most inputs, 
\begin{table}[!hbpt]

\caption{Performance comparison of solver with edge-centric ILDU0 vs. baseline implementation.}
\label{tab:eceval}
\centering
\huge
%\scalebox{0.45}
\resizebox{\columnwidth}{!}
{
\begin{tabular}{|l|ll|lll|}
\hline
               & \multicolumn{2}{l|}{Baseline OPM solver}                                                                      & \multicolumn{3}{c|}{Solver with edge-centric fused ILDU0}                                                                                                                                                                                \\ \hline
matrix         & \multicolumn{1}{l|}{iterations} & \begin{tabular}[c]{@{}l@{}}solver \\ runtime\\ on GPU\\ (sec.)\end{tabular} & \multicolumn{1}{l|}{iterations} & \multicolumn{1}{l|}{\begin{tabular}[c]{@{}l@{}}total overhead:\\ partitioning + \\ DAG generation\\ (sec.)\end{tabular}} & \begin{tabular}[c]{@{}l@{}}solver \\ runtime\\ on GPU\\ (sec.)\end{tabular} \\ \hline
laplacian1     & \multicolumn{1}{l|}{424}        & 7.15                                                                        & \multicolumn{1}{l|}{701}        & \multicolumn{1}{l|}{0.38}                                                                                                & 2.61                                                                        \\ \hline
laplacian2     & \multicolumn{1}{l|}{693}        & 26.27                                                                       & \multicolumn{1}{l|}{975}        & \multicolumn{1}{l|}{0.78}                                                                                                & 7.3                                                                         \\ \hline
parabolic\_fem & \multicolumn{1}{l|}{1631}       & 3.03                                                                        & \multicolumn{1}{l|}{2191}       & \multicolumn{1}{l|}{0.18}                                                                                                & 0.79                                                                        \\ \hline
spe10          & \multicolumn{1}{l|}{2157}       & 6.63                                                                        & \multicolumn{1}{l|}{3663}       & \multicolumn{1}{l|}{0.57}                                                                                                & 2.21                                                                        \\ \hline
rhd            & \multicolumn{1}{l|}{902}        & 5.64                                                                        & \multicolumn{1}{l|}{1183}       & \multicolumn{1}{l|}{1.13}                                                                                                & 1.28                                                                        \\ \hline
rhd-3T         & \multicolumn{1}{l|}{597}        & 10.18                                                                       & \multicolumn{1}{l|}{925}        & \multicolumn{1}{l|}{5.13}                                                                                                & 3.12                                                                        \\ \hline
oil            & \multicolumn{1}{l|}{62}         & 7.03                                                                        & \multicolumn{1}{l|}{151}        & \multicolumn{1}{l|}{25.72}                                                                                               & 2.22                                                                        \\ \hline
\end{tabular}
}
\end{table}
\begin{table}[!htbp]

\caption{Performance comparison of solver with vertex-centric ILDU0 vs. baseline implementation.}
\label{tab:vceval}
\centering
\huge
%\scalebox{0.45}
\resizebox{\columnwidth}{!}
{
\begin{tabular}{|l|ll|lll|}
\hline
               & \multicolumn{2}{l|}{Baseline OPM solver}                                                                      & \multicolumn{3}{c|}{Solver with vertex-centric fused ILDU0}                                                                                                                                                                              \\ \hline
matrix         & \multicolumn{1}{l|}{iterations} & \begin{tabular}[c]{@{}l@{}}solver \\ runtime\\ on GPU\\ (sec.)\end{tabular} & \multicolumn{1}{l|}{iterations} & \multicolumn{1}{l|}{\begin{tabular}[c]{@{}l@{}}total overhead:\\ partitioning + \\ DAG generation\\ (sec.)\end{tabular}} & \begin{tabular}[c]{@{}l@{}}solver \\ runtime\\ on GPU\\ (sec.)\end{tabular} \\ \hline
laplacian1     & \multicolumn{1}{l|}{424}        & 7.15                                                                        & \multicolumn{1}{l|}{715}        & \multicolumn{1}{l|}{0.38}                                                                                                & 3.17                                                                        \\ \hline
laplacian2     & \multicolumn{1}{l|}{693}        & 26.27                                                                       & \multicolumn{1}{l|}{923}        & \multicolumn{1}{l|}{0.78}                                                                                                & 8.08                                                                        \\ \hline
parabolic\_fem & \multicolumn{1}{l|}{1631}       & 3.03                                                                        & \multicolumn{1}{l|}{2369}       & \multicolumn{1}{l|}{0.17}                                                                                                & 0.98                                                                        \\ \hline
spe10          & \multicolumn{1}{l|}{2157}       & 6.63                                                                        & \multicolumn{1}{l|}{3659}       & \multicolumn{1}{l|}{0.57}                                                                                                & 3.26                                                                        \\ \hline
rhd            & \multicolumn{1}{l|}{902}        & 5.64                                                                        & \multicolumn{1}{l|}{1401}       & \multicolumn{1}{l|}{1.13}                                                                                                & 1.81                                                                        \\ \hline
rhd-3T         & \multicolumn{1}{l|}{597}        & 10.18                                                                       & \multicolumn{1}{l|}{953}        & \multicolumn{1}{l|}{5.12}                                                                                                & 3.91                                                                        \\ \hline
oil            & \multicolumn{1}{l|}{62}         & 7.03                                                                        & \multicolumn{1}{l|}{151}        & \multicolumn{1}{l|}{25.72}                                                                                               & 2.97                                                                        \\ \hline
\end{tabular}
}

\end{table}
As shown in Figure~\ref{fig:overallsolver}, without accounting for domain decomposition and other preprocessing overhead, our approach outperforms the baseline for all matrices. With overhead, our edge-centric kernel outperforms the baseline for all inputs except \texttt{oil} dataset which has partitioning overhead greater than the runtime of a single invocation of the solver. We emphasize performance improvement without accounting overhead for domain decomposition and other preprocessing steps for two primary reasons:

\begin{enumerate}[leftmargin=*]
  
  \item  \texttt{METIS} accounts for over 90\% of the overhead associated with domain decomposition in real-world graphs (e.g., \textit{rhd-3T} and \textit{oil}). This overhead can be further reduced using the parallel graph partitioning tools such as \texttt{ParMETIS}~\cite{parmetis}.

  \item More importantly, in many applications, the BiCGSTAB solver is repeatedly invoked on the same sparsity pattern, such as in nonlinear solvers, where the outer loop can call the iterative solver over 1000 times~\cite{nonlinear_iteration_count}. In such cases, the one-time overhead—comparable to the runtime of a few solver invocations—becomes insignificant as the accumulated performance improvements over multiple runs bring overall improvements closer to those observed without overhead.
\end{enumerate}

%\atharva{todo: mention tolerance in the tables}
%\atharva{todo: remove zero as the starting point for the residual plots
%Fix the pltos for parabolic_fem and spe10}
%\atharva{citation for multiple BiCGSTAB invocations}

\noindent Figure~\ref{fig:resplots} shows the residual at the end of each iteration for three variants of the solver: 
(1) BiCGSTAB solver with \texttt{rocSPARSE} triangular solves~\cite{rocsparse} without domain decomposition;
(2) solver with domain decomposition and vertex-centric fused ILDU0 implementation for DAG traversal, as explained in \S\ref{sssec:vc-dag}; and
(3) solver with domain decomposition and the edge-centric fused ILDU0 kernel for triangular solves, as explained in \S\ref{sssec:ec-dag}.

%\atharv{mention that variation in iteration count is the symptom /consequence of non-determinsitic atomic write order}
We observe variation in the iteration count for convergence of the solver when using the edge-centric kernels. However, the variation in iteration count was limited to $\pm$10\% of the iteration count of the vertex-centric variant, which computes triangular solves in a deterministic order.
Overall, the solver with the edge-centric ILDU0 kernel outperformed the baseline implementation by a factor of $3.2\times$ (geometric mean).
%despite requiring a $1.6\times$ more iterations to converge.

\subsection{Multi-GPU performance of ILDU0}
\begin{figure}[h]
%\vspace{-10}
\centering
\includegraphics[width=\columnwidth]{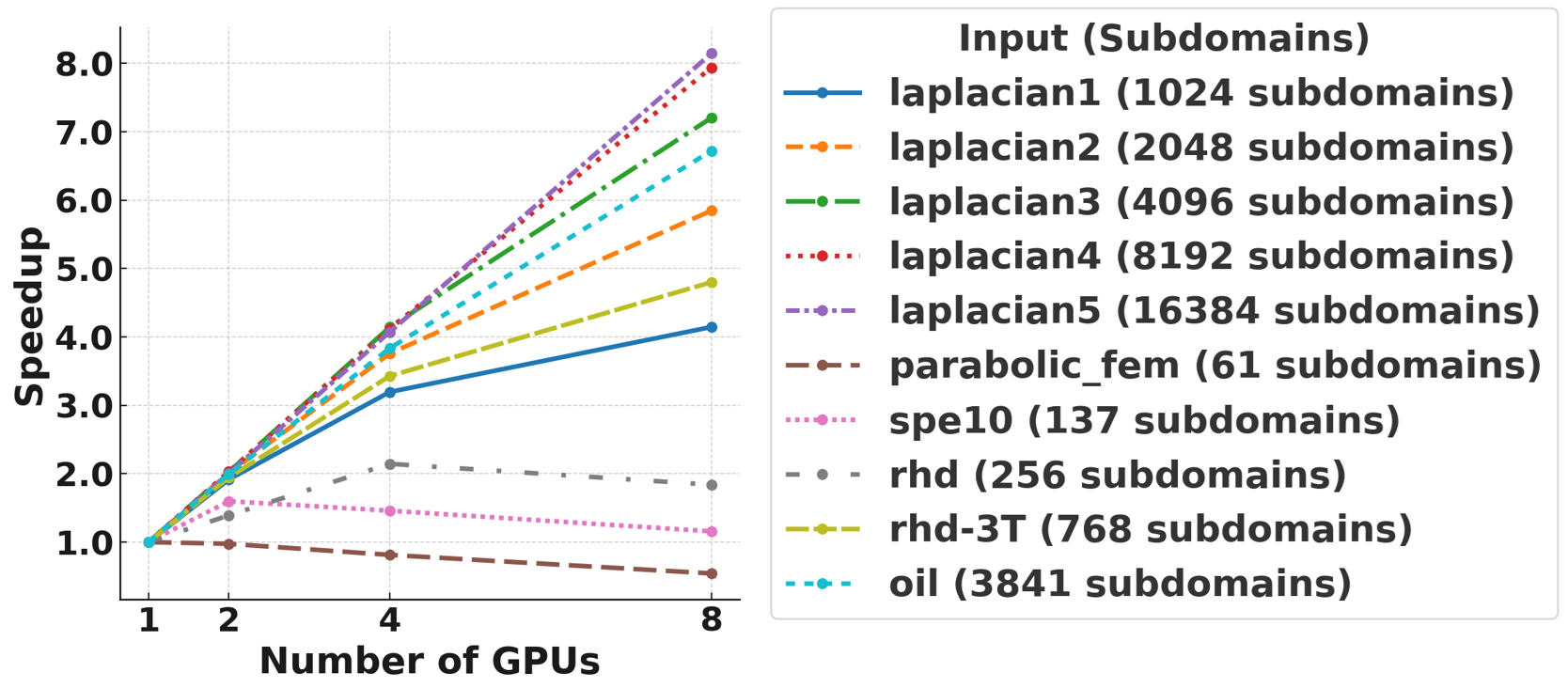}
\caption{Speedup of multi-GPU implementation of dag\_ec\_ILDU0\_fused over single-GPU performance on AMD Instinct\textsuperscript{TM} MI210.}
\label{fig:mgpusp}
\vspace{-10pt}
\end{figure}

%application}
We evaluate the multi-GPU performance of our fastest preconditioner application kernel, \texttt{dag\_ec\_ILDU0\_fused}, on up to 8 AMD Instinct MI210 GPUs. In addition to the inputs listed in Table~\ref{tab:inputs}, we include three additional block-sparse Laplacian discretizations: \textbf{laplacian3} (grid size $256 \times 256 \times 128$, approximately 8 million block rows), \textbf{laplacian4} ($256 \times 256 \times 256$, approximately 16 million block rows), and \textbf{laplacian5} ($256 \times 512 \times 256$, approximately 32 million block rows).
We use \texttt{RCCL} (ROCm Communication Collectives Library) to distribute the ILDU0 computation across GPUs. Each AMD Instinct MI210 GPU contains 104 compute units (CUs), and subdomains are mapped directly to CUs. As a result, no inter-GPU communication is needed during the execution of the multi-GPU kernel. 
 For CSR-format inputs, the subdomain size is fixed at 8192 rows. For BSR-format Laplacians, we use subdomains of 2048 blocks rows (each block of size $3 \times 3$), resulting in 6144 individual rows per subdomain.
Figure~\ref{fig:mgpusp} shows the speedup of all inputs relative to single-GPU performance.
%This enables near-linear scaling with the number of GPUs as long as (1) the total number of subdomains is evenly divisible across GPUs, and (2) each GPU is assigned more subdomains than the number of CUs.

For smaller problems where the total number of subdomains is less than the number of CUs---even on a single GPU---we observe little or no speedup. For example, \textbf{parabolic\_fem} has only 61 subdomains, which underutilizes a single GPU. When distributed across multiple GPUs, and the overhead from multi-GPU kernel launches and synchronization, results in a net performance loss.
 For larger matrices such as \textbf{laplacian3}, \textbf{laplacian4}, \textbf{laplacian5}, and \textbf{oil}, we observe near-linear speedup, as expected.

\section{Related Work}
\label{sec:related}
We categorize related work into three parts: (1) triangular solve algorithms on GPUs, (2) GPU-based BiCGSTAB solvers, and (3) domain decomposition techniques for iterative solvers.

\subsection{Parallel triangular solves}
%Parallel triangular solves are broadly classified into \emph{synchronization-free} and \emph{DAG-based} approaches. Synchronization-free methods rely on busy-wait loops, where threads stall until dependencies are resolved. DAG-based methods construct a dependency graph during an analysis phase and traverse it in a solution phase, typically using edge- or vertex-centric strategies. While DAG methods expose more parallelism, they introduce nontrivial overhead during DAG construction.

Liu et al.~\cite{liu1} propose a spin-loop-based method that avoids expensive DAG preprocessing by using row-wise nonzero counts and on-chip scratchpad memory. Their method achieves a 2.14$\times$ speedup over vendor library implementation. Liu et al.~\cite{Liu2} extend their previous work~\cite{liu1} by introducing an adaptive scheme designed to efficiently handle multiple right-hand sides in triangular solves. 
Hu et al.~\cite{hu} present AG-SpTRSV, an automated framework for optimizing sparse triangular solves through graph transformation, heuristic scheduling, and execution scheme selection. %AG-SpTRSV eliminates manual tuning and generates high-performance GPU kernels.

\texttt{rocSPARSE}~\cite{rocsparse} is the sparse linear algebra library for the AMD platform. \texttt{rocSPARSE} implements synchronization-free triangular solves with wavefronts assigned to matrix rows, and with wavefront threads processing nonzeros after dependency resolution. \texttt{rocSPARSE} supports both CSR and BSR formats. We use \texttt{rocSPARSE} triangular solves as our baseline.
Naumov~\cite{Naumov} proposes a DAG-based method with a one-time DAG generation followed by triangular solve. This method achieves up to 2$\times$ speedups over CPU MKL when applied to ILU and Cholesky preconditioners on NVIDIA GPUs.
Helal et al.~\cite{helal} optimize DAG-based solves via adaptive task aggregation (ATA) and sorted eager task (SET) scheduling, reporting 2.2$\times$--3.7$\times$ mean speedups over
%the 
existing DAG 
%execution 
approaches.
%and up to 100$\times$ speedups for large, wide DAGs with long critical paths.
Xie et al.~\cite{xie} develop a multi-GPU triangular solver using NVSHMEM for fine-grained dependency tracking and task-pool execution. Their method improves load balancing and scalability by reducing interconnect contention.

\subsection{BiConjugate Gradient Stabilized 
%Method 
%\textsf{(BiCGSTAB)}}
\small\textsf{(BiCGSTAB)}}
BiCGSTAB~\cite{vandervost-bicgstab} is a Krylov subspace method suited for unsymmetric or numerically unstable systems~\cite{saad}. Ahuja et al.~\cite{rbicgstab} present BiCGSTAB with Krylov subspace recycling for sequences of related systems.
Yamazaki et al.~\cite{yamazaki} accelerate BiCGSTAB on GPU clusters using variable-size batched kernels, achieving $4\times$ speedup over single-node performance by addressing irregular computation and communication bottlenecks.
A CUDA-based BiCGSTAB implementation~\cite{cubicgstab}, using \texttt{cuSPARSE} and \texttt{cuBLAS}, shows $2\times$ speedup over MKL-based CPU solvers. 
% Performance gains stem from optimized sparse matrix-vector products and triangular solves, especially with multiple right-hand sides.
The Open Porous Media (OPM) project~\cite{opm} provides an AMD GPU implementation using \texttt{rocSPARSE}~\cite{rocsparse} and \texttt{rocBLAS}~\cite{rocblas}, with ILU0-based preconditioning. %We use this implementation as our baseline.

\subsection{Domain decomposition for sparse linear systems}
Domain decomposition methods apply the divide-and-conquer principle to solve PDEs in 2D and 3D domains~\cite{saad}. Anzt et al.~\cite{anzt} show that approximate triangular solves, despite requiring more iterations, can outperform exact methods due to faster preconditioner application. Hong et al.~\cite{hong2} propose an adaptive domain decomposition strategy for multi-GPU systems, achieving up to $6.6\times$ speedup over a single GPU 
%performance 
via load balancing, locality-aware clustering, and overlapping computation with data transfer.

\section{Future Work}
Future direction for this work include reordering matrices post-decomposition to improve data locality, extending triangular solve kernels to non-uniform subdomain sizes, evaluating the optimizations in two-stage preconditioners, and applying the decomposition strategy to other preconditioners such as ILU-k.
% Future directions for this work include reordering the matrix after domain decomposition to improve data locality and extending the triangular solve kernels to handle non-uniform subdomain sizes. A possible extension to this work can include the evaluation of the impact of our optimizations in two-stage preconditioners,
% and application of the decomposition method used in this work to other preconditioners, such as ILU-k.
%\atharva{better versions of Bicgstab , bIcgstab -l}
%We plan to leverage spectral and other partitioning methods to reduce the number of nonzeros dropped after domain decomposition.
Another area of exploration is a coscheduled CPU-GPU implementation for triangular solves, where a subset of subdomains is assigned to the CPU while the GPU processes the remaining subdomains.

\label{sec:future}

%\input{text/bc}
%\vspace{-10}
\section{Conclusion}
\label{sec:conclusion}
This work has explored the efficacy of GPU-centric optimizations for sparse triangular solves on matrices with non-overlapping subdomains. The fine-grained domain decomposition approach presented in this work generates subdomains that can be independently processed by GPU compute units. Each subdomain is sized such that the corresponding vector data fits entirely in shared memory, enabling fast data access during triangular solves. 
Key optimizations for sparse triangular solves—such as edge-centric traversal on per-subdomain directed acyclic graphs (DAGs), shared memory usage, and kernel fusion—deliver a geometric mean speedup of $10.7\times$ over \texttt{rocSPARSE} triangular solve kernels. %We further analyze performance gains using hardware metrics such as memory bandwidth and vector ALU utilization, showing consistently higher values than \texttt{rocSPARSE}.

When evaluating 
%with 
the biconjugate gradient stabilized (BiCGSTAB) solver on both synthetic and real-world matrices, our approach achieves a $3.2\times$ geometric mean speedup over the \texttt{rocSPARSE}-based baseline on a single AMD Instinct\textsuperscript{TM} MI210 GPU, despite requiring $1.6\times$ more iterations to converge. Additionally, on a multi-GPU system with 8 AMD Instinct MI210 GPUs, our method achieves near-linear scaling relative to single GPU 
%performance 
for problems large enough to saturate all GPUs.

% This work has explored the efficacy of GPU-centric optimizations for triangular solves on matrices with non-overlapping subdomains.
% %impact of domain decomposition on the performance and  convergence of iterative solvers
% %We use the biconjugate gradient stabilized method (BiCGSTAB) as a case study to evaluate the impact of domain decomposition on the performance and convergence of the iterative solver. 
% Domain decomposition and ILU0 factorization generates a preconditioner with non-overlapping subdomains that can be processed independently. Optimizations such as DAG-based solves, assigning subdomains to thread blocks, using shared memory for vector data, and kernel fusion achieve a geometric mean speedup of $10.7\times$ over \texttt{rocSPARSE} triangular solves.
% We also analyze performance improvements using hardware metrics such as thread activity and memory bandwidth, demonstrating that our kernels achieve higher thread activity and bandwidth compared to \texttt{rocSPARSE} triangular solve kernels. 
% Our approach, evaluated on synthetic and real-world problems, achieves a $3.2\times$ geometric mean speedup over \texttt{rocSPARSE}-based BiCGSTAB on the MI210 GPU, despite requiring $1.6\times$ more iterations (geometric mean) to converge.

\balance

\begingroup
  \setlength{\itemsep}{0pt}    % no extra space between items
  \setlength{\parskip}{0pt}    % no extra paragraph skip either
  \bibliographystyle{siamplain}
  \bibliography{example_references}

\begin{thebibliography}{10}

\bibitem{mi210_bandwidth}
{\em {AMD MI210 Peak Bandwidth}}.
\newblock AMD,
  \url{https://www.futurewithamd.com/en/wp-content/uploads/sites/2/2023/03/20220407-MI210_Infographic.pdf}.
\newblock Accessed: 2025-01-01.

\bibitem{laplacian}
{\em {Finite difference method – Laplacian}}.
\newblock AMD,
  \url{https://gpuopen.com/learn/amd-lab-notes/amd-lab-notes-finite-difference-docs-laplacian_part1/#discretization-and-host-implementation}.
\newblock Accessed: 2025-01-01.

\bibitem{cubicgstab}
{\em {Incomplete-LU and Cholesky Preconditioned Iterative Methods Using
  cuSPARSE and cuBLAS}}.
\newblock NVIDIA,
  \url{https://docs.nvidia.com/cuda/incomplete-lu-cholesky/index.html}.
\newblock Accessed: 2025-01-01.

\bibitem{opm}
{\em {OPM Simulators}}.
\newblock OPM,
  \url{https://github.com/OPM/opm-simulators/tree/e667efe5227869b90117b75f57e2177b9683b585/opm/simulators/linalg/gpubridge/rocm}.
\newblock Accessed: 2025-01-01.

\bibitem{Naumov}
{\em {Parallel Solution of Sparse Triangular Linear Systems in the
  Preconditioned Iterative Methods on the GPU}}.
\newblock Maxim Naumov, NVIDIA,
  \url{https://research.nvidia.com/sites/default/files/pubs/2011-06_Parallel-Solution-of/nvr-2011-001.pdf}.
\newblock Accessed: 2025-01-01.

\bibitem{parmetis}
{\em {ParMETIS}}, \url{https://github.com/KarypisLab/ParMETIS}.
\newblock Accessed: 2025-01-01.

\bibitem{rocblas}
{\em {rocBLAS}}.
\newblock ROCm, \url{https://github.com/ROCm/rocBLAS}.
\newblock Accessed: 2025-01-01.

\bibitem{rocprof}
{\em {rocprof: ROCm Profiler}}.
\newblock AMD,
  \url{https://rocm.docs.amd.com/projects/rocprofiler/en/docs-5.5.1/rocprof.html}.
\newblock Accessed: 2025-01-01.

\bibitem{rocsparse}
{\em {rocSPARSE}}.
\newblock ROCm, \url{https://github.com/ROCm/rocSPARSE}.
\newblock Accessed: 2025-01-01.

\bibitem{spe10}
{\em {SPE Comparative Solution Project }}.
\newblock Society of Petroleum Engineers,
  \url{https://www.spe.org/web/csp/datasets/set02.htm/}.
\newblock Accessed: 2025-01-01.

\bibitem{zenodo}
{\em {StructMG: A Fast and Scalable Structured Multigrid Experiment Data and
  Source Codes}}, \url{https://doi.org/10.5281/zenodo.10346358}.
\newblock Accessed: 2025-01-01.

\bibitem{metisdocs}
{\em {METIS for Python}}, 2023, \url{https://metis.readthedocs.io/en/latest/}.
\newblock Accessed: 2025-01-03.

\bibitem{finance}
{\sc Y.~Achdou, O.~Bokanowski, and T.~Lelièvre}, {\em Partial Differential
  Equations in Finance}, John Wiley \& Sons, Ltd, 2012,
  \url{https://doi.org/https://doi.org/10.1002/9781118182635.efm0081},
  \url{https://onlinelibrary.wiley.com/doi/abs/10.1002/9781118182635.efm0081},
  \url{https://arxiv.org/abs/https://onlinelibrary.wiley.com/doi/pdf/10.1002/9781118182635.efm0081}.

\bibitem{rbicgstab}
{\sc K.~Ahuja, P.~Benner, E.~de~Sturler, and L.~Feng}, {\em Recycling bicgstab
  with an application to parametric model order reduction}, SIAM Journal on
  Scientific Computing, 37 (2015), pp.~S429--S446,
  \url{https://doi.org/10.1137/140972433},
  \url{https://doi.org/10.1137/140972433},
  \url{https://arxiv.org/abs/https://doi.org/10.1137/140972433}.

\bibitem{anzt}
{\sc H.~Anzt, E.~Chow, and J.~Dongarra}, {\em Iterative sparse triangular
  solves for preconditioning}, in Euro-Par 2015: Parallel Processing, J.~L.
  Tr{\"a}ff, S.~Hunold, and F.~Versaci, eds., Berlin, Heidelberg, 2015,
  Springer Berlin Heidelberg, pp.~650--661.

\bibitem{mp2}
{\sc M.~Baboulin, A.~Buttari, J.~J. Dongarra, J.~Kurzak, J.~Langou, J.~Langou,
  P.~Luszczek, and S.~Tomov}, {\em Accelerating scientific computations with
  mixed precision algorithms}, CoRR, abs/0808.2794 (2008),
  \url{http://arxiv.org/abs/0808.2794}, \url{https://arxiv.org/abs/0808.2794}.

\bibitem{mp1}
{\sc E.~Carson and N.~Khan}, {\em Mixed precision iterative refinement with
  sparse approximate inverse preconditioning}, SIAM Journal on Scientific
  Computing, 45 (2023), pp.~C131--C153,
  \url{https://doi.org/10.1137/22M1487709},
  \url{https://doi.org/10.1137/22M1487709},
  \url{https://arxiv.org/abs/https://doi.org/10.1137/22M1487709}.

\bibitem{spe10paper}
{\sc M.~A. Christie and M.~J. Blunt}, {\em Tenth spe comparative solution
  project: A comparison of upscaling techniques}, SPE Reservoir Evaluation \&
  Engineering, 4 (2001), pp.~308--317, \url{https://doi.org/10.2118/72469-PA},
  \url{https://doi.org/10.2118/72469-PA},
  \url{https://arxiv.org/abs/https://onepetro.org/REE/article-pdf/4/04/308/2586053/spe-72469-pa.pdf}.

\bibitem{suitesparse}
{\sc T.~A. Davis and Y.~Hu}, {\em The university of florida sparse matrix
  collection}, ACM Trans. Math. Softw., 38 (2011),
  \url{https://doi.org/10.1145/2049662.2049663},
  \url{https://doi.org/10.1145/2049662.2049663}.

\bibitem{space}
{\sc {Fortier, A.}, {Alibert, Y.}, {Carron, F.}, {Benz, W.}, and {Dittkrist,
  K.-M.}}, {\em Planet formation models: the interplay with the planetesimal
  disc}, A\&A, 549 (2013), p.~A44,
  \url{https://doi.org/10.1051/0004-6361/201220241},
  \url{https://doi.org/10.1051/0004-6361/201220241}.

\bibitem{helal}
{\sc A.~E. Helal, A.~M. Aji, M.~L. Chu, B.~M. Beckmann, and W.-c. Feng}, {\em
  Adaptive task aggregation for high-performance sparse solvers on gpus}, in
  2019 28th International Conference on Parallel Architectures and Compilation
  Techniques (PACT), 2019, pp.~324--336,
  \url{https://doi.org/10.1109/PACT.2019.00033}.

\bibitem{hong2}
{\sc S.~Hong, G.~Jang, and W.-K. Jeong}, {\em Mg-fim: A multi-gpu fast
  iterative method using adaptive domain decomposition}, SIAM Journal on
  Scientific Computing, 44 (2022), pp.~C54--C76,
  \url{https://doi.org/10.1137/21M1414644},
  \url{https://doi.org/10.1137/21M1414644},
  \url{https://arxiv.org/abs/https://doi.org/10.1137/21M1414644}.

\bibitem{hu}
{\sc Z.~Hu, J.~Sun, Z.~Li, and G.~Sun}, {\em Ag-sptrsv: An automatic framework
  to optimize sparse triangular solve on gpus}, ACM Trans. Archit. Code Optim.,
  21 (2024), \url{https://doi.org/10.1145/3674911},
  \url{https://doi.org/10.1145/3674911}.

\bibitem{edge_v_node}
{\sc Y.~Jia, V.~Lu, J.~Hoberock, M.~Garland, and J.~C. Hart}, {\em Chapter 2 -
  edge v. node parallelism for graph centrality metrics}, in GPU Computing Gems
  Jade Edition, W.~mei W.~Hwu, ed., Applications of GPU Computing Series,
  Morgan Kaufmann, Boston, 2012, pp.~15--28,
  \url{https://doi.org/https://doi.org/10.1016/B978-0-12-385963-1.00002-2},
  \url{https://www.sciencedirect.com/science/article/pii/B9780123859631000022}.

\bibitem{metis}
{\sc G.~Karypis and V.~Kumar}, {\em A fast and high quality multilevel scheme
  for partitioning irregular graphs}, SIAM Journal on Scientific Computing, 20
  (1998), pp.~359--392, \url{https://doi.org/10.1137/S1064827595287997},
  \url{https://doi.org/10.1137/S1064827595287997},
  \url{https://arxiv.org/abs/https://doi.org/10.1137/S1064827595287997}.

\bibitem{suitesparseweb}
{\sc S.~P. Kolodziej, M.~Aznaveh, M.~Bullock, J.~David, T.~A. Davis,
  M.~Henderson, Y.~Hu, and R.~Sandstrom}, {\em The suitesparse matrix
  collection website interface}, Journal of Open Source Software, 4 (2019),
  p.~1244, \url{https://doi.org/10.21105/joss.01244},
  \url{https://doi.org/10.21105/joss.01244}.

\bibitem{liu1}
{\sc W.~Liu, A.~Li, J.~Hogg, I.~S. Duff, and B.~Vinter}, {\em A
  synchronization-free algorithm for parallel sparse triangular solves}, in
  Proceedings of the 22nd International Conference on Euro-Par 2016: Parallel
  Processing - Volume 9833, Berlin, Heidelberg, 2016, Springer-Verlag,
  p.~617–630, \url{https://doi.org/10.1007/978-3-319-43659-3_45},
  \url{https://doi.org/10.1007/978-3-319-43659-3_45}.

\bibitem{Liu2}
{\sc W.~Liu, A.~Li, J.~D. Hogg, I.~S. Duff, and B.~Vinter}, {\em Fast
  synchronization-free algorithms for parallel sparse triangular solves with
  multiple right-hand sides}, Concurrency and Computation: Practice and
  Experience, 29 (2017), \url{https://doi.org/10.1002/cpe.4244},
  \url{https://www.osti.gov/biblio/1557091}.

\bibitem{parks_and_de_sturler}
{\sc M.~L. Parks, E.~de~Sturler, G.~Mackey, D.~D. Johnson, and S.~Maiti}, {\em
  Recycling krylov subspaces for sequences of linear systems}, SIAM Journal on
  Scientific Computing, 28 (2006), pp.~1651--1674,
  \url{https://doi.org/10.1137/040607277},
  \url{https://doi.org/10.1137/040607277},
  \url{https://arxiv.org/abs/https://doi.org/10.1137/040607277}.

\bibitem{saad}
{\sc Y.~Saad}, {\em Iterative Methods for Sparse Linear Systems}, Society for
  Industrial and Applied Mathematics, second~ed., 2003,
  \url{https://doi.org/10.1137/1.9780898718003},
  \url{https://epubs.siam.org/doi/abs/10.1137/1.9780898718003},
  \url{https://arxiv.org/abs/https://epubs.siam.org/doi/pdf/10.1137/1.9780898718003}.

\bibitem{ddbook}
{\sc B.~F. Smith, P.~E. Bj\o{}rstad, and W.~D. Gropp}, {\em Domain
  decomposition: parallel multilevel methods for elliptic partial differential
  equations}, Cambridge University Press, USA, 1996.

\bibitem{nonlinear_iteration_count}
{\sc T.~Spenke, N.~Delaissé, J.~Degroote, and N.~Hosters}, {\em On the number
  of subproblem iterations per coupling step in partitioned fluid-structure
  interaction simulations}, 2023, \url{https://arxiv.org/abs/2303.08513},
  \url{https://arxiv.org/abs/2303.08513}.

\bibitem{vandervost-bicgstab}
{\sc H.~A. van~der Vorst}, {\em Bi-cgstab: A fast and smoothly converging
  variant of bi-cg for the solution of nonsymmetric linear systems}, SIAM
  Journal on Scientific and Statistical Computing, 13 (1992), pp.~631--644,
  \url{https://doi.org/10.1137/0913035}, \url{https://doi.org/10.1137/0913035},
  \url{https://arxiv.org/abs/https://doi.org/10.1137/0913035}.

\bibitem{xie}
{\sc C.~Xie, J.~Chen, J.~Firoz, J.~Li, S.~L. Song, K.~Barker, M.~Raugas, and
  A.~Li}, {\em Fast and scalable sparse triangular solver for multi-gpu based
  hpc architectures}, in Proceedings of the 50th International Conference on
  Parallel Processing, ICPP '21, New York, NY, USA, 2021, Association for
  Computing Machinery, \url{https://doi.org/10.1145/3472456.3472478},
  \url{https://doi.org/10.1145/3472456.3472478}.

\bibitem{yamazaki}
{\sc I.~Yamazaki, A.~Abdelfattah, A.~Ida, S.~Ohshima, S.~Tomov, R.~Yokota, and
  J.~Dongarra}, {\em Performance of hierarchical-matrix bicgstab solver on gpu
  clusters}, 05 2018, pp.~930--939,
  \url{https://doi.org/10.1109/IPDPS.2018.00102}.

\bibitem{zong}
{\sc Y.~Zong, X.~Wang, H.~Huang, C.~Zhang, X.~Xu, J.~Sun, B.~Yan, Q.~Wang,
  S.~Li, Z.~Ding, and W.~Xue}, {\em Poster: Structmg: A fast and scalable
  structured multigrid}, in Proceedings of the 29th ACM SIGPLAN Annual
  Symposium on Principles and Practice of Parallel Programming, PPoPP '24, New
  York, NY, USA, 2024, Association for Computing Machinery, p.~478–480,
  \url{https://doi.org/10.1145/3627535.3638482},
  \url{https://doi.org/10.1145/3627535.3638482}.

\end{thebibliography}
\endgroup

%\bibliographystyle{siamplain}
%\bibliography{example_references}
\end{document}